\documentclass[11pt]{article}%
\usepackage{amssymb}
\usepackage{amsfonts}
\usepackage{amsmath}
\usepackage{graphicx}%
\providecommand{\U}[1]{\protect\rule{.1in}{.1in}}
\newtheorem{theorem}{Theorem}

\begin{document}

\title{Exact solution of the Kermack and McKendrick SIR differential equations}
\author{Piet Van Mieghem\thanks{Faculty of Electrical Engineering, Mathematics and
Computer Science, P.O Box 5031, 2600 GA Delft, The Netherlands; \emph{email}:
P.F.A.VanMieghem@tudelft.nl. }}
\date{Delft University of Technology\\
1 September 2020}
\maketitle

\begin{abstract}
Several exact expansions as well as lower and upperbounds of the Kermack and
McKendrick SIR equations are presented.

\end{abstract}

\section{SIR governing equations}

\label{sec_SIR_governing_equations}In their seminal paper
\cite{Kermack_McKendrick1927}, Kermack and McKendrick derive the differential
equations for SIR epidemics in a homogeneous population (i.e. complete graph)
with constant infection rate $\beta$ and curing rate $\delta$
\begin{equation}%
\begin{array}
[c]{ccc}%
\frac{dx}{dt}=-\beta xy & \frac{dy}{dt}=\beta xy-\delta y & \frac{dz}%
{dt}=\delta y
\end{array}
\label{Kermack_McKendrick_SIR_dvgl}%
\end{equation}
where $x,y,z$ denotes the number of susceptible, infected and removed items in
a fixed population of size $N=x+y+z$. The set
(\ref{Kermack_McKendrick_SIR_dvgl}) is a special case of the general
Kermack-McKendrick theory for constant rates. The Kermack-McKendrick
differential equations with constant rates $\beta$ and $\delta$ in
(\ref{Kermack_McKendrick_SIR_dvgl}) describe the basic SIR\ model for a
disease without re-infections and appear in nearly each book and course on
epidemics (see e.g.
\cite{Anderson_May,Diekmann_Heesterbeek_Britton_boek2012,Kiss_Miller_Simon2016,Daley}%
). Even today in Corona times, predictions and first order estimates of
infected individuals are based on the SIR equations
(\ref{Kermack_McKendrick_SIR_dvgl}).

Here, we present exact solutions, which, at the best of our knowledge, have
not yet appeared inspite of the fundamental role of the SIR differential
equation (\ref{Kermack_McKendrick_SIR_dvgl}) in the theory of epidemics.
Numerous approximate solutions of (\ref{Kermack_McKendrick_SIR_dvgl}) exist
(see e.g. \cite{Harko2014,Barlow2020}) and the first approximation is
presented by Kermack and McKendrick \cite{Kermack_McKendrick1927}, which is
here revisited and generalized. Tedious mathematical derivations are placed in Appendices.

As usual in SIS epidemics, we denote the effective infection rate $\tau
=\frac{\beta}{\delta}$, which is equal to the basic reproduction number
$R_{0}$. A key observation of Kermack and McKendrick
\cite{Kermack_McKendrick1927} is that
\[
\frac{dx}{dz}=-\tau x
\]
whose solution is $\log\frac{x\left(  t\right)  }{x_{0}}=-\tau z\left(
t\right)  $, because initially there are no removed, $z\left(  0\right)  =0$,
while $x_{0}=x\left(  0\right)  $ is the initial number of susceptible items.
Writing $y=N-x-z$ in the last SIR differential equation in
(\ref{Kermack_McKendrick_SIR_dvgl}) and introducing $x=x_{0}e^{-\tau z}$
yields%
\begin{equation}
\frac{dz}{dt}=\delta\left(  N-x_{0}e^{-\tau z}-z\right)
\label{Kermack_McKendrick_exact_dvgl_SIR_removed_items}%
\end{equation}
Hence, the set of differential equations in (\ref{Kermack_McKendrick_SIR_dvgl}%
) is equivalent to%
\[%
\begin{array}
[c]{ccc}%
x=x_{0}e^{-\tau z} & y=N-x-z & \frac{dz}{dt}=\delta\left(  N-x_{0}e^{-\tau
z}-z\right)
\end{array}
\]
where only one differential equation
(\ref{Kermack_McKendrick_exact_dvgl_SIR_removed_items}) remains.

Kermack and McKendrick \cite{Kermack_McKendrick1927} integrate
(\ref{Kermack_McKendrick_exact_dvgl_SIR_removed_items}) with the scaled time
$t^{\ast}=\delta t$, taken into account that $z\left(  0\right)  =0$, and
present the exact result%
\begin{equation}
t^{\ast}=\int_{0}^{z}\frac{du}{N-x_{0}e^{-\tau u}-u}
\label{Kermack_McKendrick_integral_SIR_removed_items}%
\end{equation}
If the effective infection rate $\tau$ is a function of time $t$, then the
differential equation (\ref{Kermack_McKendrick_exact_dvgl_SIR_removed_items})
cannot be directly integrated anymore. In other words, the confinement to
constant rates greatly simplifies the analysis of the SIR differential
equations. This paper mainly concentrates on the differential
(\ref{Kermack_McKendrick_exact_dvgl_SIR_removed_items}) and the integral
(\ref{Kermack_McKendrick_integral_SIR_removed_items}).

The parameter~$N$ is eliminated if we define the fraction of susceptible items
by $\xi=\frac{x}{N}$, of infected by $\eta=\frac{y}{N}$ and of removed by
$\zeta=\frac{z}{N}$ so that%
\[
\xi+\eta+\zeta=1
\]
but the initial conditions with a zero recovered fraction, $\zeta_{0}=0$, obey%
\[
\xi_{0}=1-\eta_{0}%
\]
The integral (\ref{Kermack_McKendrick_integral_SIR_removed_items}) for the
scaled time becomes%
\[
t^{\ast}=\frac{1}{N}\int_{0}^{N\zeta}\frac{du}{1-\xi_{0}e^{-\tau u}-\frac
{1}{N}u}%
\]
We define the normalized effective infection rate by $\theta=N\tau$ and,
expect from SIS epidemics \cite{PVM_MSIS_star_PRE2012} on the complete graph,
that the epidemic threshold $\tau_{c}\approx\frac{1}{N}$ and $\theta
_{c}\approx1$. In other words, the generalization to networks would be
$\theta_{G}=\frac{\tau}{\tau_{c}}$, where $\tau_{c}$ is the epidemic threshold
for SIR spread in a graph $G$. Let $w=\frac{u}{N}$, then we arrive at the
(scaled) time $t^{\ast}=\delta t$, measured in units of the average curing
time $\frac{1}{\delta}$, as a function of the fraction $\zeta$ of removed
items in a homogeneous population or complete graph,%
\begin{equation}
t^{\ast}=t_{\theta}^{\ast}\left(  \zeta\right)  =\int_{0}^{\zeta}\frac
{dw}{1-\xi_{0}e^{-\theta w}-w}
\label{Kermack_McKendrick_integral_SIR_removed_fractions}%
\end{equation}
The last differential equation in (\ref{Kermack_McKendrick_SIR_dvgl}) in terms
of fractions, $\frac{d\zeta}{dt^{\ast}}=\eta$, indicates that the fraction
$\zeta$ of removed strictly increases with time $t^{\ast}$ until the fraction
of infected $\eta$ equals zero, where $\zeta$ attains a maximum $\zeta_{\max}%
$. Since the fraction of infected $\eta=1-\xi_{0}e^{-\theta\zeta}-\zeta\geq0$,
it follows that $1-\zeta\geq\xi_{0}e^{-\theta\zeta}$ and equality when
$\eta=0$ corresponds to the maximal fraction $\zeta_{\max}$ of removed items.
At $w=\zeta_{\max}$, the denominator of the integral in
(\ref{Kermack_McKendrick_integral_SIR_removed_fractions}) is zero and the
corresponding time $t_{\theta}^{\ast}\left(  \zeta_{\max}\right)  $ is
obtained after infinitely long time. We require physically that the fraction
of removed $\zeta\in\lbrack0,\zeta_{\max})$. The maximal fraction $\zeta
_{\max}$ is expressed in terms of the Lambert function \cite{Corless1996} in
(\ref{max_fraction_removed_Lambert}) in Appendix \ref{sec_Lambert_function}.
Fig. \ref{Fig_maxremovedfractionSIR} plots the maximum removed fraction
$\zeta_{\max}$ computed by (\ref{max_fraction_removed_Lambert}) as a function
of the initial fraction $\xi_{0}$ of susceptible for various normalized
effective infection rates $\theta$, starting from $\theta=0.2$ up to
$\theta=2.0$ in steps of $0.2$.
\begin{figure}
[h]
\begin{center}
\includegraphics[
height=7.2862cm,
width=12.1144cm
]%
{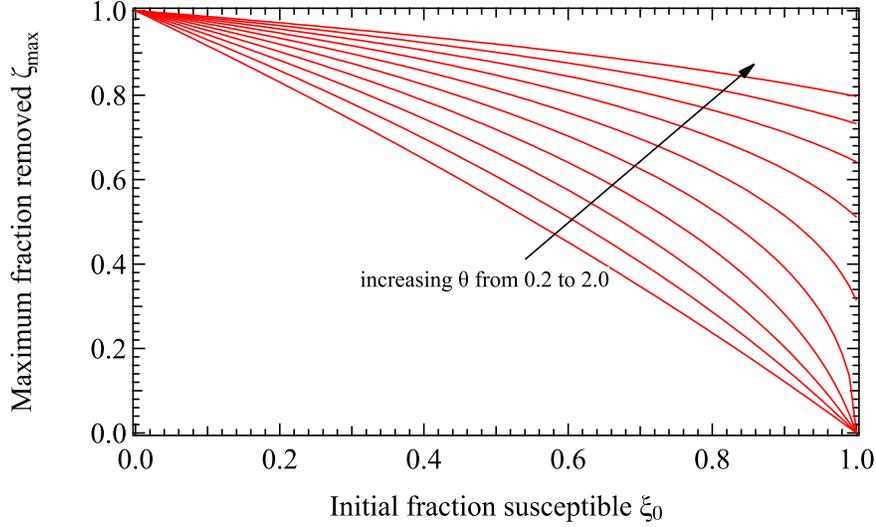}%
\caption{The maximum fraction $\zeta_{\max}$ of removed items in an SIR
epidemics versus the initial fraction $\xi_{0}$ of susceptible, for various
normalized effective infection rates $\theta$.}%
\label{Fig_maxremovedfractionSIR}%
\end{center}
\end{figure}

\section{Solution of the SIR governing equations}

\label{sec_SIR_governing_equations_solution}Formally, the exact solution
(\ref{Kermack_McKendrick_integral_SIR_removed_items}) of the Kermack and
McKendrick SIR differential equation
(\ref{Kermack_McKendrick_exact_dvgl_SIR_removed_items}) expresses the scaled
time $t^{\ast}=H\left(  \zeta\right)  $ in terms of the fraction $\zeta$ of
removed items, where the integral is%
\[
H\left(  w\right)  =\int_{0}^{w}\frac{du}{h\left(  u\right)  }%
\]
with $h\left(  u\right)  =1-\xi_{0}e^{-\theta u}-u$. Since $h\left(  u\right)
\geq0$, the integral $H\left(  w\right)  $ is increasing in $w\geq0$.
Moreover, fractions are contained in $\left[  0,1\right]  $ and $h\left(
u\right)  \leq1$, which implies that $H\left(  w\right)  \geq w$.
Clearly\footnote{If $t^{\ast}=H\left(  z\right)  $ is continuous and strictly
increasing from $t_{1}^{\ast}$ to $t_{2}^{\ast}$ as $z$ increases from $z_{1}$
to $z_{2}$, then there is a unique inverse function $z=H^{-1}\left(  t^{\ast
}\right)  $, which is also continuous and strictly increasing from $z_{1}$ to
$z_{2}$ as $t^{\ast}$ increases from $t_{1}^{\ast}$ to $t_{2}^{\ast}$. This
theorem is proved in \cite[p. 206]{Hardy_pure_math}.}, there exists an inverse
function $H^{-1}$ so that $\zeta=H^{-1}\left(  t^{\ast}\right)  $ and
$t^{\ast}=H\left(  \zeta\right)  $, similarly as $t=\arcsin y$, where $\arcsin
y=\int_{0}^{y}\frac{du}{\sqrt{1-u^{2}}}$ and $y=\sin t$. From the key property
of inverse functions%
\[
H\left(  H^{-1}\left(  t^{\ast}\right)  \right)  =t^{\ast}%
\]
differentiation yields%
\[
\frac{dH^{-1}\left(  t^{\ast}\right)  }{dt^{\ast}}=\frac{1}{\left.
\frac{dH\left(  x\right)  }{dx}\right\vert _{x=H^{-1}\left(  t^{\ast}\right)
}}=h\left(  H^{-1}\left(  t^{\ast}\right)  \right)
\]
which is nothing else than the differential equation
(\ref{Kermack_McKendrick_exact_dvgl_SIR_removed_items}).

Since the integral (\ref{Kermack_McKendrick_integral_SIR_removed_items}) is
not analytically known, Kermack and McKendrick approximate $e^{-\theta\zeta
}=1-\theta\zeta+\frac{1}{2}\theta^{2}\zeta^{2}+O\left(  \zeta^{3}\right)  $ up
to third order in (\ref{Kermack_McKendrick_integral_SIR_removed_fractions}) to
obtain%
\[
\frac{d\zeta}{dt^{\ast}}=1-\xi_{0}+\left(  \xi_{0}\theta-1\right)  \zeta
-\frac{\xi_{0}\theta^{2}}{2}\zeta^{2}%
\]
which is a Riccati differential equation%
\begin{equation}
\frac{dw}{dt}=aw-bw^{2}-c \label{dvgl_SIS_logistic}%
\end{equation}
whose solution is%
\begin{equation}
w\left(  t\right)  =\frac{a}{2b}+\frac{\Upsilon}{2b}\tanh\left(  \frac{t}%
{2}\Upsilon+\text{arctanh}\left(  \frac{2by_{0}-a}{\Upsilon}\right)  \right)
\label{Riccati_y_tanh_constant_c}%
\end{equation}
where $\Upsilon=\sqrt{a^{2}-4bc}$. The solution
(\ref{Riccati_y_tanh_constant_c}) appeared already in
\cite{Kermack_McKendrick1927} and is reviewed in \cite[Sec. 2.3]{Daley}. The
Riccati differential equation (\ref{dvgl_SIS_logistic}) is directly integrated
as
\[
t_{3}=\int_{w_{0}}^{w}\frac{du}{au-bu^{2}-c}%
\]
which equals (after rewriting $au-bu^{2}-c=\frac{\Upsilon^{2}}{4b}\left\{
1-\left(  \frac{2b}{\Upsilon}\left(  u-\frac{a}{2b}\right)  \right)
^{2}\right\}  $)
\[
t_{3}=\int_{w_{0}}^{w}\frac{du}{\frac{\Upsilon^{2}}{4b}\left\{  1-\left(
\frac{2b}{\Upsilon}\left(  u-\frac{a}{2b}\right)  \right)  ^{2}\right\}
}=\frac{2}{\Upsilon}\left.  \text{arctanh}\frac{2b}{\Upsilon}\left(
u-\frac{a}{2b}\right)  \right\vert _{w_{0}}^{w}%
\]
Inversion (i.e. solving for $w$) leads to (\ref{Riccati_y_tanh_constant_c}).
Inserting $a=\left(  \xi_{0}\theta-1\right)  $, $b=\frac{\xi_{0}\theta^{2}}%
{2}$ and $c=-\left(  1-\xi_{0}\right)  $ provides us with the approximation
$t_{3}$ for the time $t^{\ast}$ as function of the fraction $\zeta$ of removed
items in the population,%
\begin{equation}
t_{3}=\frac{2}{\Upsilon}\left\{  \text{arctanh}\left(  \frac{\xi_{0}\theta
^{2}\zeta-\left(  \xi_{0}\theta-1\right)  }{\Upsilon}\right)  +\text{arctanh}%
\frac{\left(  \xi_{0}\theta-1\right)  }{\Upsilon}\right\}
\label{upper_bound_time_3order_fractions}%
\end{equation}
with%
\[
\Upsilon=\sqrt{\left(  \xi_{0}\theta-1\right)  ^{2}+2\theta^{2}\xi_{0}\left(
1-\xi_{0}\right)  }%
\]
Since \cite[p. 103]{PVM_PAComplexNetsCUP}%
\[
e^{-\theta\zeta}<1-\theta\zeta+\frac{1}{2}\theta^{2}\zeta^{2}%
\]
we conclude that (\ref{upper_bound_time_3order_fractions}) derived from the
third order approximation in $e^{-\theta\zeta}$ upper bounds the correct
time,
\[
t_{\theta}^{\ast}\left(  \zeta\right)  <t_{3}%
\]
Consequently, the inverse relation deduced from
(\ref{upper_bound_time_3order_fractions}) indicates that%
\[
\frac{2}{\Upsilon}\left\{  \text{arctanh}\left(  \frac{\xi_{0}\theta^{2}%
\zeta-\left(  \xi_{0}\theta-1\right)  }{\Upsilon}\right)  +\text{arctanh}%
\frac{\left(  \xi_{0}\theta-1\right)  }{\Upsilon}\right\}  >t_{\theta}^{\ast
}\left(  \zeta\right)
\]
and%
\begin{equation}
\zeta\left(  t^{\ast}\right)  >\frac{1}{\xi_{0}\theta^{2}}\left\{  \left(
\xi_{0}\theta-1\right)  +\Upsilon\tanh\left(  \frac{\Upsilon}{2}t^{\ast
}-\text{arctanh}\frac{\left(  \xi_{0}\theta-1\right)  }{\Upsilon}\right)
\right\}  \label{zeta_tanh_lower_bound}%
\end{equation}
In other words, the \textquotedblleft tanh\textquotedblright-approximation
underestimates the fraction of removed items. Equivalently, the conservation
law $\xi+\eta+\zeta=1$ implies that the \textquotedblleft
tanh\textquotedblright-approximation overestimates the fraction $\eta$ of
infection items, as demonstrated earlier for SIS epidemics
\cite{PVM_tanh_formula_prevalence},\cite{PVM_bounds_SIS_prevalence}.

\subsection{The \textquotedblleft tanh\textquotedblright-approximation for the
average path length in small-world graphs}

The \textquotedblleft tanh\textquotedblright-approximation also appears in an
approximate, but ingenious computation in \cite{Newman_PRL2000} of the average
path length in small-world graphs \cite{WattsStrogatzNature1998}. The
Watts-Strogatz small-world graph $G_{WS}\left(  p_{r},k,N\right)  $ has $N$
nodes regularly placed and consecutively numbered on a ring. Each node $i$ has
$2k$ links connected to its direct neighbors $i-k,i-k+1,\ldots,i-1,i+1,\ldots
,i+k$ and the basic law of the degree $\sum_{j=1}^{N}d_{j}=2L$ then tells us
that the number of links $L=kN$. Each end point of a link has probability
$p_{r}$ to be rewired to a random node; in total, there are $s=p_{r}kN$
rewired links, called shortcuts. Newman \emph{et al}. \cite{Newman_PRL2000}
consider a continuous version of the Watts-Strogatz small-world graph
$G_{WS}\left(  p_{r},k,N\right)  $, where the one-dimensional ring lattice is
treated as a continuum and shortcuts are assumed to have zero length. The
neighborhood $b\left(  r\right)  $ of segment length $r$ around a random point
(node) on the circle consists of the set of points that can be reached by
following paths of length $r$ or less on the graph $G_{WS}\left(
p_{r},k,N\right)  $. The fraction $q\left(  r\right)  $ of points that belongs
to a neighborhood $b\left(  r\right)  $ follows from \cite{Newman_PRL2000} as%
\begin{equation}
r=-\frac{1}{4k^{2}p_{r}}\int_{0}^{q}\frac{dv}{v^{2}-v-\frac{1}{2Nkp_{r}}}
\label{r_versus_q_Newman_small_world}%
\end{equation}
The average path length or hopcount (i.e. number of links in the shortest
path) is deduced in \cite{Newman_PRL2000} as $E\left[  H\right]  =-\int
_{0}^{1}rdq$. After introducing (\ref{r_versus_q_Newman_small_world}) and
performing a partial integration, we find the basic result of Newman \emph{et
al.}%
\[
E\left[  H\right]  =\frac{1}{4k^{2}p_{r}}\int_{0}^{1}\frac{\left(  1-v\right)
dv}{v^{2}-v-\frac{1}{2Nkp_{r}}}=\frac{1}{2k^{2}p_{r}}\frac{1}{\sqrt{1+\frac
{2}{Nkp_{r}}}}\arctan\frac{1}{\sqrt{1+\frac{2}{Nkp_{r}}}}%
\]
The scaled approximate time $t_{3}$ satisfies%
\[
t_{3}=-\frac{1}{\frac{\xi_{0}\theta^{2}}{2}}\int_{0}^{\zeta}\frac{du}%
{u^{2}-\frac{2}{\xi_{0}\theta^{2}}\left(  \xi_{0}\theta-1\right)
u-\frac{2\left(  1-\xi_{0}\right)  }{\xi_{0}\theta^{2}}}%
\]
and suggests the analogy between a segment length $r$ versus scaled time
$t_{3}$ and between the fraction $q$ of points that belongs to a (random)
neighborhood $b\left(  r\right)  $ versus the fraction $\zeta$ of removed
items in an SIR epidemics.

\subsection{Partial fraction expansion}

Here, we present a formal generalization to any order $m$ in $O\left(
\zeta^{m}\right)  $. First up to $O\left(  \zeta^{4}\right)  $, the bound for
any real $\theta$ \cite[p. 103]{PVM_PAComplexNetsCUP}
\[
e^{-\theta\zeta}>1-\theta\zeta+\frac{1}{2}\theta^{2}\zeta^{2}-\frac{1}%
{6}\theta^{3}\zeta^{3}%
\]
illustrates that increasing $m$ alternatively provides lower and upper bounds.
Introduced into Kermack and McKendrick differential equation
(\ref{Kermack_McKendrick_integral_SIR_removed_fractions}) shows\footnote{This
differential equation with a third order polynomial resembles that of
Weierstrass's elliptic $\mathcal{P}\left(  z\right)  $ function \cite[p.
247]{Sansone},
\[
\left(  \frac{d\mathcal{P}\left(  z\right)  }{dz}\right)  ^{2}=4\mathcal{P}%
^{3}\left(  z\right)  -g_{2}\mathcal{P}\left(  z\right)  +g_{3}%
\]
} that%
\[
\frac{d\zeta}{dt^{\ast}}<\left(  1-\xi_{0}\right)  +\left(  \xi_{0}%
\theta-1\right)  \zeta-\frac{\xi_{0}\theta^{2}}{2}\zeta^{2}+\frac{1}{6}\xi
_{0}\theta^{3}\zeta^{3}%
\]
The third order polynomial $p_{3}\left(  \zeta\right)  $ at the right-hand
side can be factored as
\[
p_{3}\left(  \zeta\right)  =A\left(  \zeta-\zeta_{1}\right)  \left(
\zeta-\zeta_{2}\right)  \left(  \zeta-\zeta_{3}\right)
\]
where $A=\frac{1}{6}\xi_{0}\theta^{3}$. The zeros $\zeta_{1},\zeta_{2}$ and
$\zeta_{3}$ can be analytically expressed by Cardano's formulas for the cubic.
Thus, we have%
\[
\frac{d\zeta}{dt^{\ast}}<A\left(  \zeta-\zeta_{1}\right)  \left(  \zeta
-\zeta_{2}\right)  \left(  \zeta-\zeta_{3}\right)
\]
from which%
\[
\frac{d\zeta}{\left(  \zeta-\zeta_{1}\right)  \left(  \zeta-\zeta_{2}\right)
\left(  \zeta-\zeta_{3}\right)  }<Adt^{\ast}%
\]
After integration and partial fraction expansion (provided all zeros
$\zeta_{1}$, $\zeta_{2}$ and $\zeta_{3}$ are different)%
\[
\frac{1}{\left(  \zeta-\zeta_{1}\right)  \left(  \zeta-\zeta_{2}\right)
\left(  \zeta-\zeta_{3}\right)  }=\frac{a_{1}}{\left(  \zeta-\zeta_{1}\right)
}+\frac{a_{2}}{\left(  \zeta-\zeta_{2}\right)  }+\frac{a_{3}}{\left(
\zeta-\zeta_{3}\right)  }%
\]
we find, with $a_{1}=\frac{1}{\left(  \zeta_{1}-\zeta_{2}\right)  \left(
\zeta_{1}-\zeta_{3}\right)  }$, $a_{2}=\frac{1}{\left(  \zeta_{2}-\zeta
_{1}\right)  \left(  \zeta_{2}-\zeta_{3}\right)  }$ and $a_{3}=\frac
{1}{\left(  \zeta_{3}-\zeta_{1}\right)  \left(  \zeta_{3}-\zeta_{2}\right)  }%
$,
\[
\int_{0}^{\zeta}\frac{a_{1}dw}{\left(  w-\zeta_{1}\right)  }+\int_{0}^{\zeta
}\frac{a_{2}dw}{\left(  w-\zeta_{2}\right)  }+\int_{0}^{\zeta}\frac{a_{3}%
dw}{\left(  w-\zeta_{3}\right)  }<At^{\ast}%
\]
Hence, we arrive at%
\[
\log\left(  \frac{\zeta-\zeta_{1}}{\zeta_{1}}\right)  ^{a_{1}}\left(
\frac{\zeta-\zeta_{2}}{\zeta_{2}}\right)  ^{a_{2}}\left(  \frac{\zeta
-\zeta_{3}}{\zeta_{3}}\right)  ^{a_{3}}<At^{\ast}%
\]
from which the lower bound follows%
\[
\left(  \frac{\zeta-\zeta_{1}}{\zeta_{1}}\right)  ^{a_{1}}\left(  \frac
{\zeta-\zeta_{2}}{\zeta_{2}}\right)  ^{a_{2}}\left(  \frac{\zeta-\zeta_{3}%
}{\zeta_{3}}\right)  ^{a_{3}}<e^{At^{\ast}}%
\]
In general, we cannot solve $\zeta$ from this inequality. After increasing the
order to $O\left(  \zeta^{5}\right)  $, the quartic with zeros $\omega
_{1},\omega_{2},\omega_{3}$ and $\omega_{4}$ leads to the upper bound%
\[
\left(  \frac{\zeta-\omega_{1}}{\omega_{1}}\right)  ^{\alpha_{1}}\left(
\frac{\zeta-\omega_{2}}{\omega_{2}}\right)  ^{\alpha_{2}}\left(  \frac
{\zeta-\omega_{3}}{\omega_{3}}\right)  ^{\alpha_{3}}\left(  \frac{\zeta
-\omega_{4}}{\omega_{4}}\right)  ^{\alpha_{3}}>e^{At^{\ast}}%
\]
Formally, the partial fraction method can be extended to any polynomial and to
the exact case itself, as shown below.

Cauchy's integral theorem \cite{Titchmarshfunctions} states that%

\[
\frac{1}{1-\xi_{0}e^{-\theta w}-w}=\frac{1}{2\pi i}\int_{C\left(  w\right)
}\frac{1}{1-\xi_{0}e^{-\theta z}-z}\frac{dz}{\left(  z-w\right)  }%
\]
where the contour $C\left(  w\right)  $ encloses in counter-clockwise sense a
region around the point $z=w$, where the integrand is analytic. Since%
\[
\lim_{r\rightarrow\infty}\frac{1}{1-\xi_{0}e^{-\theta re^{i\omega}%
}-re^{i\omega}}=0
\]
for any angle $\omega$, the integrand vanishes for $\left\vert z\right\vert
\rightarrow\infty$ and we can deform the contour to enclose the entire complex
plane without the point $z=w$, in clockwise sense,
\[
\frac{1}{2\pi i}\int_{C\left(  w\right)  }\frac{1}{1-\xi_{0}e^{-\theta z}%
-z}\frac{dz}{\left(  z-w\right)  }=-\frac{1}{2\pi i}\int_{C\backslash\left\{
w\right\}  }\frac{1}{1-\xi_{0}e^{-\theta z}-z}\frac{dz}{\left(  z-w\right)  }%
\]
The function $\frac{1}{1-\xi_{0}e^{-\theta z}-z}$ has poles at the zeros of
$1-\xi_{0}e^{-\theta z}-z$, where only $0\leq\arg z<2\pi$ is enclosed by the
contour. The simple zero $\widetilde{z}$ obeys $1-\widetilde{z}=\xi
_{0}e^{-\theta\widetilde{z}}$, which, as shown in Section
\ref{sec_Lambert_function}, can be transformed to $qe^{-q}=a$ with
$a=\theta\xi_{0}e^{-\theta}\in\left[  0,\xi_{0}\right]  $. Section
\ref{sec_complex_zeros_qexp(-q)=a} illustrates that there are infinitely many
complex zeros $\left\{  \widetilde{z}_{k}\right\}  _{k\geq0}$, whose precise
form can only be computed numerically. Cauchy's residue theorem tells us that%
\begin{align*}
\frac{1}{1-\xi_{0}e^{-\theta w}-w}  &  =\sum_{\widetilde{z}_{k}}\frac
{1}{w-\widetilde{z}_{k}}\lim_{z\rightarrow\widetilde{z}_{k}}\frac
{z-\widetilde{z}_{k}}{1-\xi_{0}e^{-\theta z}-z}=\sum_{\widetilde{z}_{k}}%
\frac{1}{w-\widetilde{z}_{k}}\frac{1}{\xi_{0}\theta e^{-\theta\widetilde
{z}_{k}}-1}\\
&  =\sum_{\widetilde{z}_{k}}\frac{1}{w-\widetilde{z}_{k}}\frac{1}%
{\theta-1-\theta\widetilde{z}_{k}}%
\end{align*}
This result is the partial fraction expansion of $\frac{1}{1-\xi_{0}e^{-\theta
w}-w}$ in terms of its complex zeros. The scaled time in
(\ref{Kermack_McKendrick_integral_SIR_removed_fractions}) becomes%
\begin{align*}
t^{\ast}  &  =\int_{0}^{\zeta}\frac{dw}{1-\xi_{0}e^{-\theta w}-w}\\
&  =\sum_{\widetilde{z}_{k}}\int_{0}^{\zeta}\frac{dw}{w-\widetilde{z}_{k}%
}\frac{1}{\theta-1-\theta\widetilde{z}_{k}}=\sum_{\widetilde{z}_{k}}%
\log\left(  \frac{\widetilde{z}_{k}-\zeta}{\widetilde{z}_{k}}\right)  \frac
{1}{\theta-1-\theta\widetilde{z}_{k}}%
\end{align*}
and%
\[
e^{t^{\ast}}=\prod_{\widetilde{z}_{k}}\left(  1-\frac{\zeta}{\widetilde{z}%
_{k}}\right)  ^{\frac{1}{\theta-1-\theta\widetilde{z}_{k}}}%
\]

Section \ref{sec_complex_zeros_qexp(-q)=a} shows that there is only one real
zero $\zeta_{\max}$ specified in (\ref{max_fraction_removed_Lambert}), while
all others zeros,%
\[
\widetilde{z}_{k}=1-\frac{x_{k}+iy_{k}}{\theta}=\frac{\theta-x_{k}-iy_{k}%
}{\theta}%
\]
are complex conjugate (with $x_{k}>0$), where $q=x_{k}+iy_{k}$ satisfies
$qe^{-q}=a>0$. Thus, for real $w$, we obtain
\begin{align*}
\frac{1}{1-\xi_{0}e^{-\theta w}-w}  &  =\frac{1}{w-\zeta_{\max}}\frac
{1}{\theta-1-\theta\zeta_{\max}}+2\theta\sum_{y_{k}>0}\operatorname{Re}\left(
\frac{1}{\theta w+x_{k}-\theta+iy_{k}}\frac{1}{x_{k}-1+iy_{k}}\right) \\
&  =\frac{1}{w-\zeta_{\max}}\frac{1}{\theta-1-\theta\zeta_{\max}}+2\theta
\sum_{y_{k}>0}\frac{\left(  \theta w+x_{k}-\theta\right)  \left(
x_{k}-1\right)  -y_{k}^{2}}{\left(  \left(  \theta w+x_{k}-\theta\right)
^{2}+y_{k}^{2}\right)  \left(  x_{k}-1\right)  ^{2}+y_{k}^{2}}%
\end{align*}
and analogously, after some tedious calculations,
\begin{equation}
t^{\ast}=\frac{\log\left(  1-\frac{\zeta}{\zeta_{\max}}\right)  }%
{\theta-1-\theta\zeta_{\max}}+2\sum_{y_{k}>0}\frac{\log\left\vert
1+\frac{2x_{k}-\theta\left(  2-\zeta\right)  }{\left(  \theta-x_{k}\right)
^{2}+y_{k}^{2}}\theta\zeta\right\vert \left(  x_{k}-1\right)  +y_{k}%
\arctan\frac{\theta\zeta y_{k}}{\left(  \theta-x_{k}\right)  ^{2}+y_{k}%
^{2}+\theta\zeta\left(  x_{k}-\theta\right)  }}{\left(  x_{k}-1\right)
^{2}+y_{k}^{2}} \label{t_scaled_complex_zeros_qexp(-q)=a}%
\end{equation}
where $x_{k}^{2}+y_{k}^{2}=a^{2}e^{2x_{k}}$ grows exponentially fast. Because
the complex zeros $\widetilde{z}_{k}=1-\frac{x_{k}+iy_{k}}{\theta}$ can only
be numerically computed, we do not further investigate this novel approach
(\ref{t_scaled_complex_zeros_qexp(-q)=a}), but concentrate on series
expansions in Section \ref{sec_series_scaled_time}.

\section{Bounds on the scaled time $t^{\ast}$}

\label{sec_bounds}Before turning to an exact series expansion of the scaled
time $t^{\ast}$ in Section \ref{sec_series_scaled_time}, we present a set of
different bounds.

The integral (\ref{Kermack_McKendrick_integral_SIR_removed_fractions}) is
analytically computable in two extreme limits of the normalized effective
infection rate $\theta$. First, if $\theta\rightarrow\infty$, then%
\[
t_{\theta\rightarrow\infty}^{\ast}\left(  \zeta\right)  =\lim_{\theta
\rightarrow\infty}\int_{0}^{\zeta}\frac{dw}{1-\xi_{0}e^{-\theta w}-w}=\int
_{0}^{\zeta}\frac{dw}{1-w}%
\]
and%
\[
t_{\theta\rightarrow\infty}^{\ast}\left(  \zeta\right)  =-\log\left(
1-\zeta\right)
\]
Thus, if the infectiousness is unlimitedly strong $\theta\rightarrow\infty$,
then the removed fraction is $\zeta_{\left\{  \theta\rightarrow\infty\right\}
}\left(  t^{\ast}\right)  =\left(  1-e^{-t^{\ast}}\right)  $. The other
extremal case for $\theta\rightarrow0$ is%
\[
t_{\theta\rightarrow0}^{\ast}\left(  \zeta\right)  =\lim_{\theta\rightarrow
0}\int_{0}^{\zeta}\frac{dw}{1-\xi_{0}e^{-\theta w}-w}=\int_{0}^{\zeta}%
\frac{dw}{1-\xi_{0}-w}%
\]
and%
\[
t_{\theta\rightarrow0}^{\ast}\left(  \zeta\right)  =-\log\left(  1-\frac
{\zeta}{1-\xi_{0}}\right)
\]
Thus, if the infectious power is absent $\theta\rightarrow0$, then the removed
fraction is $\zeta_{\left\{  \theta\rightarrow0\right\}  }\left(  t^{\ast
}\right)  =\left(  1-\xi_{0}\right)  \left(  1-e^{-t^{\ast}}\right)  $. In
summary, the fraction $\zeta_{\tau}\left(  t^{\ast}\right)  $ of removed items
as a function of the scaled time $t^{\ast}$ is bounded by%
\[
\left(  1-\xi_{0}\right)  \left(  1-e^{-t^{\ast}}\right)  \leq\zeta_{\tau
}\left(  t^{\ast}\right)  \leq\left(  1-e^{-t^{\ast}}\right)
\]
Alternatively, the scaled time $t^{\ast}=t_{\theta}^{\ast}\left(
\zeta\right)  =\int_{0}^{\zeta}\frac{dw}{\left(  1-\xi_{0}e^{-\theta
w}\right)  -w}$ is bounded by
\begin{equation}
-\log\left(  1-\frac{\zeta}{1-\xi_{0}}\right)  \leq t_{\theta}^{\ast}\left(
\zeta\right)  \leq-\log\left(  1-\zeta\right)
\label{bounds_scaled_time_extreme_theta_s}%
\end{equation}
Since $1-\frac{\zeta}{1-\xi_{0}}=\frac{1-\xi_{0}-\zeta}{1-\xi_{0}}$, while the
fraction of infected $\eta=1-\xi_{0}e^{-\theta\zeta}-\zeta$ at any time, the
above inequality suggests a reasonable estimate,%
\begin{equation}
t_{\theta}^{\ast}\left(  \zeta\right)  >-\log\left(  1-\frac{\zeta}{1-\xi
_{0}e^{-\theta\zeta}}\right)  \label{lower_bound_guess}%
\end{equation}
Numerical computations indicate that the right-hand side is a (strict) lower
bound for $t_{\theta}^{\ast}\left(  \zeta\right)  $.

Since the fraction of removed $\zeta\in\left[  0,1\right]  $, it holds that
$1-\xi_{0}e^{-\theta w}-w\leq1-\xi_{0}e^{-\theta w}$ and the integral
(\ref{Kermack_McKendrick_integral_SIR_removed_fractions}) is bounded as%
\[
t_{\theta}^{\ast}\left(  \zeta\right)  \geq\int_{0}^{\zeta}\frac{dw}{1-\xi
_{0}e^{-\theta w}}=\frac{1}{\theta}\log\left(  \frac{e^{\theta\zeta}-\xi_{0}%
}{1-\xi_{0}}\right)
\]
We rewrite $\frac{1}{\theta}\log\left(  \frac{e^{\theta\zeta}-\xi_{0}}%
{1-\xi_{0}}\right)  =\zeta-\frac{1}{\theta}\log\left(  1-\frac{\xi_{0}-\xi
_{0}e^{-\theta\zeta}}{1-\xi_{0}e^{-\theta\zeta}}\right)  $, where $\frac
{\xi_{0}-\xi_{0}e^{-\theta\zeta}}{1-\xi_{0}e^{-\theta\zeta}}\leq1$, and find%
\[
t_{\theta}^{\ast}\left(  \zeta\right)  \geq\zeta-\frac{1}{\theta}\log\left(
1-\frac{\xi_{0}-\xi_{0}e^{-\theta\zeta}}{1-\xi_{0}e^{-\theta\zeta}}\right)
\geq\zeta
\]
where the last inequality follows directly from
(\ref{Kermack_McKendrick_integral_SIR_removed_fractions}), because $1-\xi
_{0}e^{-\theta w}-w\leq1$ for $w\in\left[  0,\zeta\right]  $. The scaled time
$t_{\theta}^{\ast}\left(  \zeta\right)  $ is always larger than the fraction
of removed at that time. The above suggests us to rewrite
(\ref{Kermack_McKendrick_integral_SIR_removed_fractions}) with%
\[
\frac{1}{1-\xi_{0}e^{-\theta w}-w}=\frac{1}{\left(  1-\xi_{0}e^{-\theta
w}\right)  \left(  1-\frac{w}{1-\xi_{0}e^{-\theta w}}\right)  }%
\]
Since the fraction of infected $\eta=1-\xi_{0}e^{-\theta\zeta}-\zeta\geq0$ and
$1-\xi_{0}e^{-\theta w}-w\geq0$ for any $w\in\left[  0,\zeta\right]  $ -- the
integration parameter $w$ physically represents the fraction of removed at a
time $t^{\prime}\in\lbrack0,t]$ --, the last inequality is equivalent to
$1\geq\frac{w}{1-\xi_{0}e^{-\theta w}}$. Geometric series expansion then
yields%
\[
\frac{1}{1-\xi_{0}e^{-\theta w}-w}=\sum_{k=0}^{\infty}\frac{w^{k}}{\left(
1-\xi_{0}e^{-\theta w}\right)  ^{k+1}}=\frac{1}{1-\xi_{0}e^{-\theta w}}%
+\sum_{k=1}^{\infty}\frac{w^{k}}{\left(  1-\xi_{0}e^{-\theta w}\right)
^{k+1}}%
\]
Hence\footnote{Any Taylor series can be integrated within its region of
convergence, because it represents then an analytic function in the complex
plane.}, the integral (\ref{Kermack_McKendrick_integral_SIR_removed_fractions}%
) equals%
\begin{align*}
t_{\theta}^{\ast}\left(  \zeta\right)   &  =\int_{0}^{\zeta}\frac{dw}%
{1-\xi_{0}e^{-\theta w}-w}=\int_{0}^{\zeta}\frac{dw}{1-\xi_{0}e^{-\theta w}%
}+\sum_{k=1}^{\infty}\int_{0}^{\zeta}\frac{w^{k}dw}{\left(  1-\xi
_{0}e^{-\theta w}\right)  ^{k+1}}\\
&  =\frac{1}{\theta}\log\left(  \frac{e^{\theta\zeta}-\xi_{0}}{1-\xi_{0}%
}\right)  +\sum_{k=1}^{\infty}\int_{0}^{\zeta}\frac{w^{k}dw}{\left(  1-\xi
_{0}e^{-\theta w}\right)  ^{k+1}}%
\end{align*}
but none of the positive terms in the \thinspace$k$-sum is analytically
integrable. However, the rather trivial bounds%
\[
\frac{1}{\left(  1-\xi_{0}e^{-\theta\zeta}\right)  ^{k+1}}\int_{0}^{\zeta
}w^{k}dw\leq\int_{0}^{\zeta}\frac{w^{k}dw}{\left(  1-\xi_{0}e^{-\theta
w}\right)  ^{k+1}}\leq\frac{1}{\left(  1-\xi_{0}\right)  ^{k+1}}\int
_{0}^{\zeta}w^{k}dw
\]
lead to%
\[
\sum_{k=1}^{\infty}\frac{1}{k+1}\left(  \frac{\zeta}{1-\xi_{0}e^{-\theta\zeta
}}\right)  ^{k+1}\leq\sum_{k=1}^{\infty}\int_{0}^{\zeta}\frac{w^{k}dw}{\left(
1-\xi_{0}e^{-\theta w}\right)  ^{k+1}}\leq\sum_{k=1}^{\infty}\frac{1}%
{k+1}\left(  \frac{\zeta}{1-\xi_{0}}\right)  ^{k+1}%
\]
With $\sum_{k=1}^{\infty}\frac{x^{k+1}}{k+1}=-\log\left(  1-x\right)  -x$, we
thus obtain the bounds for $T_{\theta}^{\ast}\left(  \zeta\right)  =$
$t_{\theta}^{\ast}\left(  \zeta\right)  -\frac{1}{\theta}\log\left(
\frac{e^{\theta\zeta}-\xi_{0}}{1-\xi_{0}}\right)  $,
\begin{equation}
-\log\left(  1-\frac{\zeta}{1-\xi_{0}e^{-\theta\zeta}}\right)  -\frac{\zeta
}{1-\xi_{0}e^{-\theta\zeta}}\leq T_{\theta}^{\ast}\left(  \zeta\right)
\leq-\log\left(  1-\frac{\zeta}{1-\xi_{0}}\right)  -\frac{\zeta}{1-\xi_{0}}
\label{bounds_scaled_time_geometric_series}%
\end{equation}
The bounds in (\ref{bounds_scaled_time_geometric_series}) are clearly sharper
than the bounds in (\ref{bounds_scaled_time_extreme_theta_s}), which are
limiting cases in the normalized effective infection rate $\theta$. Instead of
bounding the integral as here, an exact series approach is presented in
Theorem \ref{theorem_Taylor_series_SIR_time}.

Numerical evaluations indicate that the scaled time $t^{\ast}=t_{\theta}%
^{\ast}\left(  \zeta\right)  $ is accurately bounded as%
\begin{equation}
\frac{1}{\theta}\log\left(  \frac{e^{\theta\zeta}-\xi_{0}}{1-\xi_{0}}\right)
-\frac{\zeta}{1-\xi_{0}e^{-\theta\zeta}}-\log\left(  1-\frac{\zeta}{1-\xi
_{0}e^{-\theta\zeta}}\right)  <t_{\theta}^{\ast}\left(  \zeta\right)  <t_{3}
\label{bounds_scaled_time}%
\end{equation}
In other words, the best lower bound deduced here appears in
(\ref{bounds_scaled_time_geometric_series}) and the best upper bound is
$t_{3}$ specified in (\ref{upper_bound_time_3order_fractions}). Finally, we
observe that the last sum in the complex zeros expansion
(\ref{t_scaled_complex_zeros_qexp(-q)=a}) only contains positive terms. Hence,
in terms of the maximum fraction $\zeta_{\max}$ of removed items specified in
(\ref{max_fraction_removed_Lambert}) in Appendix \ref{sec_Lambert_function},
we find another lower bound%
\[
t_{\theta}^{\ast}\left(  \zeta\right)  >\frac{\log\left(  1-\frac{\zeta}%
{\zeta_{\max}}\right)  }{\theta\left(  1-\zeta_{\max}\right)  -1}%
\]
which is reasonably accurate.

\section{Series for the scaled time $t^{\ast}$ in
(\ref{Kermack_McKendrick_integral_SIR_removed_items})}

\label{sec_series_scaled_time}Our major exact result is

\begin{theorem}
\label{theorem_Taylor_series_SIR_time} In the complete graph $K_{N}$ on $N$
nodes, the SIR time $t^{\ast}=\delta t$, measured in units of the average
curing time $\frac{1}{\delta}$, can be expanded in a converging series for
$\zeta<\zeta_{\max}$ specified in (\ref{max_fraction_removed_Lambert}),%
\begin{equation}
t^{\ast}=\frac{\zeta}{1-\frac{\zeta}{2}-\xi_{0}e^{-\frac{\theta\zeta}{2}}%
}\left\{  1+2\sum_{m=1}^{\infty}\left[  \sum_{k=1}^{2m}\frac{k!\sum_{j=0}%
^{k}\binom{2m}{j}\left(  \xi_{0}e^{-\frac{\theta\zeta}{2}}\right)
^{k-j}\left(  -\theta\right)  ^{2m-j}\mathcal{S}_{2m-j}^{(k-j)}}{\left(
1-\frac{\zeta}{2}-\xi_{0}e^{-\frac{\theta\zeta}{2}}\right)  ^{k}}\right]
\,\frac{\left(  \frac{\zeta}{2}\right)  ^{2m}}{\left(  2m+1\right)
!}\right\}  \label{scaled_time_converging_series_fractions}%
\end{equation}
where $\mathcal{S}_{m}^{(k)}$ is the Stirling Number of the second kind.
\end{theorem}

The proof is given in Appendix
\ref{sec_proof_Taylor_expansion_time_in_removed}. The Taylor series in
(\ref{Taylor_series_H(z)_rond_z0}) can be inverted using Lagrange series. Our
characteristic coefficients \cite[Sec. 2]{PVM_ASYM} can produce that Lagrange
series formally to any desired order term. Unfortunately, that exact Lagrange
series of $\zeta$ in terms of $t^{\ast}$ is quite involved and omitted.
Instead, we derive the Taylor series of $\zeta\left(  t^{\ast}\right)  $
around an arbitrary point $t_{0}^{\ast}$ in Section
\ref{sec_Taylor_series_diffvlg_SIR}.

All terms in the $m$-series in (\ref{scaled_time_converging_series_fractions})
are positive. Hence, summing terms up to $m\leq K$ provides a lower bound,
that is increasingly sharp for increasing $K$. However, the $k$-series in
(\ref{scaled_time_converging_series_fractions}) is alternating and causes
numerical instabilities for large $m$. In Appendix
\ref{sec_summing_Taylor_series}, we present an alternative Taylor series which
is numerically stable. Moreover, we demonstrate that the entire Taylor series
can, in principle be analytically evaluated term by term. The first split-off
of terms yields%
\begin{align}
t^{\ast}  &  =\frac{1}{1-\theta\xi_{0}e^{-\theta z_{0}}}\ln\left(  \frac
{1-\xi_{0}e^{-\theta z_{0}}\left(  1+z_{0}\theta\right)  }{1-\zeta-\xi
_{0}e^{-\theta z_{0}}\left(  1+\left(  z_{0}-\zeta\right)  \theta\right)
}\right) \nonumber\\
&  +\sum_{m=1}^{\infty}\left[  \sum_{k=1}^{m-1}\left(  \frac{\theta\xi
_{0}e^{-\theta z_{0}}-1}{\theta\left(  1-z_{0}-\xi_{0}e^{-\theta z_{0}%
}\right)  }\right)  ^{k}\sum_{j=1}^{m-k}{\binom{k}{j}}\,\left(  \frac{\xi
_{0}\theta e^{-\theta z_{0}}}{\theta\xi_{0}e^{-\theta z_{0}}-1}\right)
^{j}j!T\left(  j,m-k\right)  \right]  \,\frac{\theta^{m}z_{0}^{m+1}-\theta
^{m}\left(  z_{0}-\zeta\right)  ^{m+1}}{\left(  1-z_{0}-\xi_{0}e^{-\theta
z_{0}}\right)  \left(  m+1\right)  } \label{t*_with_log_term_chcster}%
\end{align}
The second split-off, specified by the upper-index $k=m-2$ in the $k$-sum, is%
\begin{align}
t^{\ast}  &  =\frac{1}{1-\theta\xi_{0}e^{-\theta z_{0}}}\log\left(
\frac{1-\xi_{0}e^{-\theta z_{0}}\left(  1+z_{0}\theta\right)  }{1-\zeta
-\xi_{0}e^{-\theta z_{0}}\left(  1+\left(  z_{0}-\zeta\right)  \theta\right)
}\right)  \left\{  1-\frac{\xi_{0}\theta^{2}e^{-\theta z_{0}}\left(
1-z_{0}-\xi_{0}e^{-\theta z_{0}}\right)  }{\left(  1-\theta\xi_{0}e^{-\theta
z_{0}}\right)  ^{2}}\right\} \nonumber\\
&  +\frac{1}{2}\left(  \frac{\xi_{0}\theta^{2}e^{-\theta z_{0}}}{1-\theta
\xi_{0}e^{-\theta z_{0}}}\right)  \left\{  \,\,\frac{\left(  z_{0}%
-\zeta\right)  ^{2}}{1-\zeta-\xi_{0}e^{-\theta z_{0}}\left(  1+\left(
z_{0}-\zeta\right)  \theta\right)  }-\frac{z_{0}^{2}}{1-\xi_{0}e^{-\theta
z_{0}}\left(  1+z_{0}\theta\right)  }\right\} \nonumber\\
&  +\,\frac{\zeta\xi_{0}\theta^{2}e^{-\theta z_{0}}}{\left(  1-\theta\xi
_{0}e^{-\theta z_{0}}\right)  ^{2}}\nonumber\\
&  +\sum_{m=1}^{\infty}\left[  \sum_{k=1}^{m-2}\left(  \frac{\theta\xi
_{0}e^{-\theta z_{0}}-1}{\theta\left(  1-z_{0}-\xi_{0}e^{-\theta z_{0}%
}\right)  }\right)  ^{k}\sum_{j=1}^{m-k}{\binom{k}{j}}\,\left(  \frac{\xi
_{0}\theta e^{-\theta z_{0}}}{\theta\xi_{0}e^{-\theta z_{0}}-1}\right)
^{j}j!T\left(  j,m-k\right)  \right]  \,\frac{\theta^{m}z_{0}^{m+1}-\theta
^{m}\left(  z_{0}-\zeta\right)  ^{m+1}}{\left(  1-z_{0}-\xi_{0}e^{-\theta
z_{0}}\right)  \left(  m+1\right)  } \label{t*_afgesplits_k=m-1}%
\end{align}
The third split-off with upper-index $k=m-3$ is
\begin{align}
t^{\ast}  &  =\frac{1}{1-\theta\xi_{0}e^{-\theta z_{0}}}\log\left(
\frac{1-\xi_{0}e^{-\theta z_{0}}\left(  1+z_{0}\theta\right)  }{1-\zeta
-\xi_{0}e^{-\theta z_{0}}\left(  1+\left(  z_{0}-\zeta\right)  \theta\right)
}\right) \nonumber\\
&  \times\left\{  1-\frac{\xi_{0}\theta^{2}e^{-\theta z_{0}}\left(
1-z_{0}-\xi_{0}e^{-\theta z_{0}}\right)  }{\left(  1-\theta\xi_{0}e^{-\theta
z_{0}}\right)  ^{2}}+\frac{\xi_{0}\theta^{3}e^{-\theta z_{0}}\left(
1-z_{0}-\xi_{0}e^{-\theta z_{0}}\right)  ^{2}}{2\left(  1-\theta\xi
_{0}e^{-\theta z_{0}}\right)  ^{3}}+\frac{3\xi_{0}^{2}\theta^{4}e^{-2\theta
z_{0}}\left(  1-z_{0}-\xi_{0}e^{-\theta z_{0}}\right)  ^{2}}{2\left(
1-\theta\xi_{0}e^{-\theta z_{0}}\right)  ^{4}}\right\} \nonumber\\
&  +\frac{1}{2}\left(  \frac{\xi_{0}\theta^{2}e^{-\theta z_{0}}}{1-\theta
\xi_{0}e^{-\theta z_{0}}}\right)  \left\{  \,\frac{\left(  z_{0}-\zeta\right)
^{2}}{1-\zeta-\xi_{0}e^{-\theta z_{0}}\left(  1+\left(  z_{0}-\zeta\right)
\theta\right)  }-\frac{z_{0}^{2}}{1-\xi_{0}e^{-\theta z_{0}}\left(
1+z_{0}\theta\right)  }\right\} \nonumber\\
&  \times\left\{  1-\frac{\theta\left(  1-z_{0}-\xi_{0}e^{-\theta z_{0}%
}\right)  }{3\left(  1-\theta\xi_{0}e^{-\theta z_{0}}\right)  }-\frac{3\xi
_{0}\theta^{2}e^{-\theta z_{0}}\left(  1-z_{0}-\xi_{0}e^{-\theta z_{0}%
}\right)  }{2\left(  1-\theta\xi_{0}e^{-\theta z_{0}}\right)  ^{2}}\right\}
\nonumber\\
&  +\frac{\xi_{0}^{2}\theta^{4}e^{-2\theta z_{0}}\left(  1-z_{0}-\xi
_{0}e^{-\theta z_{0}}\right)  ^{2}}{8\left(  1-\theta\xi_{0}e^{-\theta z_{0}%
}\right)  ^{3}}\left\{  \frac{\left(  z_{0}-\zeta\right)  ^{2}}{\left(
1-\zeta-\xi_{0}e^{-\theta z_{0}}\left(  1+\left(  z_{0}-\zeta\right)
\theta\right)  \right)  ^{2}}-\frac{z_{0}^{2}}{\left(  1-\xi_{0}e^{-\theta
z_{0}}\left(  1+z_{0}\theta\right)  \right)  ^{2}}\right\} \nonumber\\
&  +\,\frac{\zeta\xi_{0}\theta^{2}e^{-\theta z_{0}}}{\left(  1-\theta\xi
_{0}e^{-\theta z_{0}}\right)  ^{2}}\left\{  1-\frac{\theta\left(  1-z_{0}%
-\xi_{0}e^{-\theta z_{0}}\right)  }{2\left(  1-\theta\xi_{0}e^{-\theta z_{0}%
}\right)  }\,-\frac{3\xi_{0}\theta^{2}e^{-\theta\zeta_{0}}\left(  1-z_{0}%
-\xi_{0}e^{-\theta z_{0}}\right)  }{2\left(  1-\theta\xi_{0}e^{-\theta z_{0}%
}\right)  ^{2}}\right\} \nonumber\\
&  +\sum_{m=1}^{\infty}\left[  \sum_{k=1}^{m-3}\left(  \frac{\theta\xi
_{0}e^{-\theta z_{0}}-1}{\theta\left(  1-z_{0}-\xi_{0}e^{-\theta z_{0}%
}\right)  }\right)  ^{k}\sum_{j=1}^{m-k}{\binom{k}{j}}\,\left(  \frac{\xi
_{0}\theta e^{-\theta z_{0}}}{\theta\xi_{0}e^{-\theta z_{0}}-1}\right)
^{j}j!T\left(  j,m-k\right)  \right]  \,\frac{\theta^{m}z_{0}^{m+1}-\theta
^{m}\left(  z_{0}-\zeta\right)  ^{m+1}}{\left(  1-z_{0}-\xi_{0}e^{-\theta
z_{0}}\right)  \left(  m+1\right)  } \label{t*_afgesplits_k=m-2}%
\end{align}
When neglecting the $m$-sum in (\ref{t*_with_log_term_chcster}),
(\ref{t*_afgesplits_k=m-1}) and (\ref{t*_afgesplits_k=m-2}) increasingly
sharper lower bounds for $t^{\ast}$ are established. Although we can continue
the computations as shown in Appendix \ref{sec_summing_Taylor_series}, the
analytic terms (without $m$-sum) are already involved. Only when compared
close to divergence point where $\zeta\rightarrow\zeta_{\max}$, differences
are apparent, but for a less extreme parameter range, the best candidate
(\ref{t*_afgesplits_k=m-2}) with expansion point $z_{0}=\frac{\zeta}{2}$ is
sufficiently accurate.

\subsection{Another type of expansion}

Another application of (\ref{repeated_partial_integration_single_integral}) is
based upon%
\[
\frac{1}{1-u-\xi_{0}e^{-\theta u}}=\frac{1}{\theta\xi_{0}e^{-\theta u}-1}%
\frac{d}{du}\log\left(  1-u-\xi_{0}e^{-\theta u}\right)
\]
For $f\left(  u\right)  =\frac{d}{du}\log\left(  1-u-\xi_{0}e^{-\theta
u}\right)  $ and $g\left(  u\right)  =\frac{1}{\theta\xi_{0}e^{-\theta u}-1}$,
we obtain from (\ref{repeated_partial_integration_single_integral})
\begin{align}
\int_{0}^{\zeta}\frac{du}{1-u-\xi_{0}e^{-\theta u}}  &  =\int_{0}^{\zeta
}\left(  \sum_{k=0}^{m-1}\frac{g^{\left(  k\right)  }\left(  \zeta\right)
}{k!}\left(  u-\zeta\right)  ^{k}\right)  \frac{d}{du}\log\left(  1-u-\xi
_{0}e^{-\theta u}\right)  du\nonumber\\
&  +\frac{\left(  -1\right)  ^{m}}{\left(  m-1\right)  !}\int_{0}^{\zeta
}dx\;g^{\left(  m\right)  }\left(  x\right)  \int_{0}^{x}\left(  x-u\right)
^{m-1}\frac{d}{du}\log\left(  1-u-\xi_{0}e^{-\theta u}\right)  du
\label{SIR_integral_partial_integration_1}%
\end{align}
Partial integration of (\ref{SIR_integral_partial_integration_1}) leads after
tedious manipulations to%
\begin{align}
\int_{0}^{\zeta}\frac{du}{1-\xi_{0}e^{-\theta u}-u}  &  =\left\{  \log\left(
\frac{1-\xi_{0}e^{-\theta\zeta}-\zeta}{1-\xi_{0}}\right)  \right\}  g\left(
\zeta\right)  -\int_{0}^{\zeta}dx\;g^{\left(  1\right)  }\left(  x\right)
\log\left(  \frac{1-\xi_{0}e^{-\theta x}-x}{1-\xi_{0}}\right) \nonumber\\
&  +\log\left(  1-\xi_{0}\right)  \left\{  1_{\left\{  m>1\right\}  }\int
_{0}^{\zeta}dx\;\frac{\left(  -1\right)  ^{m}g^{\left(  m\right)  }\left(
x\right)  }{\left(  m-1\right)  !}x^{m-1}-\sum_{k=1}^{m-1}\frac{g^{\left(
k\right)  }\left(  \zeta\right)  }{k!}\left(  -\zeta\right)  ^{k}\right\}
\nonumber\\
&  -\sum_{k=0}^{m-2}\frac{g^{\left(  k+1\right)  }\left(  \zeta\right)  }%
{k!}\int_{0}^{\zeta}\left(  u-\zeta\right)  ^{k}\log\left(  1-\xi
_{0}e^{-\theta u}-u\right)  du\nonumber\\
&  +\int_{0}^{\zeta}dx\;\frac{\left(  -1\right)  ^{m}g^{\left(  m\right)
}\left(  x\right)  }{\left(  m-2\right)  !}\int_{0}^{x}\left(  x-u\right)
^{m-2}\log\left(  1-\xi_{0}e^{-\theta u}-u\right)  du
\label{SIR_integral_partial_integration_2}%
\end{align}
The first term in (\ref{SIR_integral_partial_integration_2})%
\[
t_{\theta}^{\ast}\left(  \zeta\right)  \approx\frac{\log\left(  \frac
{1-\xi_{0}e^{-\theta\zeta}-\zeta}{1-\xi_{0}}\right)  }{e^{-\theta\zeta
+\log\theta\xi_{0}}-1}%
\]
turns out to be a reasonably accurate estimate of $t^{\ast}$ for not too large
$\theta$. In fact, for $\theta\leq1$, numerical computations seem to indicate
that the above first term is a tighter lower bound than
(\ref{lower_bound_guess}).

\subsection{Time of the peak infection}

The maximum number of infected obeys $\frac{dy}{dt}=\beta xy-\delta y=0$, from
which the peak number $y_{p}=1-x_{p}-z_{p}$ of infected occurs when
$x_{p}=\frac{1}{\tau}$. Using $\log\frac{x\left(  t\right)  }{x_{0}}=-\tau
z\left(  t\right)  $, it holds that $\frac{\log x_{0}\tau}{\tau}=z_{p}$ and
the peak number of infected $y_{p}=1-\frac{1+\log x_{0}\tau}{\tau}$. Turning
to the fraction of removed $\zeta_{p}=$ $\frac{\log\xi_{0}\theta}{\theta}$ at
a maximum fraction of infected $\eta_{p}$ and using
(\ref{Kermack_McKendrick_integral_SIR_removed_fractions}) expresses the time
$t_{\text{peak}}^{\ast}=\delta t_{\text{peak}}$, expressed in units of the
average curing time $\frac{1}{\delta}$, at which the peak infection occurs
with $\theta=N\tau$ as%
\[
t_{\text{peak}}^{\ast}=\int_{0}^{\frac{\log\xi_{0}\theta}{\theta}}\frac
{dw}{1-\xi_{0}e^{-\theta w}-w}%
\]
It just remains to substitute $\zeta_{p}=$ $\frac{\log\xi_{0}\theta}{\theta}$,
$e^{-\theta\frac{\zeta_{p}}{2}}=\frac{1}{\sqrt{\xi_{0}\theta}}$ and $\xi
_{0}e^{-\theta\frac{\zeta_{p}}{2}}=\sqrt{\frac{\xi_{0}}{\theta}}$ into one of
the series (\ref{t*_with_log_term_chcster}), (\ref{t*_afgesplits_k=m-1}) and
(\ref{t*_afgesplits_k=m-2}) to find a good lower bound for $t_{\text{peak}%
}^{\ast}$.

\section{Differential equation
(\ref{Kermack_McKendrick_exact_dvgl_SIR_removed_items})}

\label{sec_Taylor_series_diffvlg_SIR}So far, we have concentrated on the
function $t^{\ast}=H\left(  \zeta\right)  $ and now we focus on $\zeta
=H^{-1}\left(  t^{\ast}\right)  $. We start a Taylor series approach and
introduce $\zeta\left(  t^{\ast}\right)  =\sum_{k=0}^{\infty}\zeta_{k}\left(
t_{0}^{\ast}\right)  \left(  t^{\ast}-t_{0}^{\ast}\right)  ^{k}$ into the
Kermack and McKendrick differential equation
(\ref{Kermack_McKendrick_exact_dvgl_SIR_removed_items}), written in
fractions,
\[
\frac{d\zeta\left(  t^{\ast}\right)  }{dt^{\ast}}=1-\xi_{0}e^{-\theta
\zeta\left(  t^{\ast}\right)  }-\zeta\left(  t^{\ast}\right)
\]
Invoking our general Taylor expansion (see Appendix \ref{sec_chc})%
\begin{equation}
e^{-\theta\;\zeta(t^{\ast})}=e^{-\theta\;\zeta_{0}\left(  t_{0}^{\ast}\right)
}\left(  1+\sum_{m=1}^{\infty}\left[  \sum_{k=1}^{m}\frac{\left(
-\theta\right)  ^{k}}{k!}\,\left.  s[k,m]\right\vert _{\zeta\left(  t\right)
}\left(  t_{0}^{\ast}\right)  \right]  \,\left(  t^{\ast}-t_{0}^{\ast}\right)
^{m}\right)  \label{expf}%
\end{equation}
where $\left.  s[k,m]\right\vert _{\zeta\left(  t\right)  }\left(  t_{0}%
^{\ast}\right)  $ is the characteristic coefficient of $\zeta\left(  t\right)
$ around $t_{0}^{\ast}$, yields%
\begin{align*}
\sum_{m=0}^{\infty}\left(  m+1\right)  \zeta_{m+1}\left(  t_{0}^{\ast}\right)
\left(  t^{\ast}-t_{0}^{\ast}\right)  ^{m}  &  =1-\xi_{0}e^{-\theta\;\zeta
_{0}\left(  t_{0}^{\ast}\right)  }-\zeta_{0}\left(  t_{0}^{\ast}\right) \\
&  -\sum_{m=1}^{\infty}\left[  \xi_{0}e^{-\theta\;\zeta_{0}\left(  t_{0}%
^{\ast}\right)  }\sum_{k=1}^{m}\frac{\left(  -\theta\right)  ^{k}}%
{k!}\,\left.  s[k,m]\right\vert _{\zeta\left(  t\right)  }\left(  t_{0}^{\ast
}\right)  +\zeta_{m}\left(  t_{0}^{\ast}\right)  \right]  \,\left(  t^{\ast
}-t_{0}^{\ast}\right)  ^{m}%
\end{align*}
Equating corresponding powers in $t^{\ast}-t_{0}^{\ast}$ results in $\zeta
_{1}\left(  t_{0}^{\ast}\right)  =1-\xi_{0}e^{-\tau\;\zeta_{0}\left(
t_{0}^{\ast}\right)  }-\zeta_{0}\left(  t_{0}^{\ast}\right)  $, which is the
differential equation at the scaled time $t_{0}^{\ast}$, and in the recursion%
\begin{equation}
\zeta_{m}\left(  t_{0}^{\ast}\right)  =-\frac{1}{m}\left(  \xi_{0}%
e^{-\theta\;\zeta_{0}\left(  t_{0}^{\ast}\right)  }\sum_{k=1}^{m-1}%
\frac{\left(  -\theta\right)  ^{k}}{k!}\,\left.  s[k,m-1]\right\vert
_{\zeta\left(  t\right)  }\left(  t_{0}^{\ast}\right)  +\zeta_{m-1}\left(
t_{0}^{\ast}\right)  \right)  \label{recursion_zeta_m}%
\end{equation}
that essentially extends the first order differential equation to all higher
orders. For example, for $m=2$ in (\ref{recursion_zeta_m}), we obtain%
\begin{align*}
\zeta_{2}\left(  t_{0}^{\ast}\right)   &  =-\frac{1}{2}\left(  1-\theta\xi
_{0}e^{-\tau\;\zeta_{0}\left(  t_{0}^{\ast}\right)  }\right)  \,\zeta
_{1}\left(  t_{0}^{\ast}\right) \\
&  =-\frac{1}{2}\left(  1-\theta\xi_{0}e^{-\tau\;\zeta_{0}\left(  t_{0}^{\ast
}\right)  }\right)  \,\left(  1-\xi_{0}e^{-\tau\;\zeta_{0}\left(  t_{0}^{\ast
}\right)  }-\zeta_{0}\left(  t_{0}^{\ast}\right)  \right)
\end{align*}
We can iterate the recursion (\ref{recursion_zeta_m}) up to any $m$. However,
the unknown $\zeta_{0}\left(  t_{0}^{\ast}\right)  =\zeta\left(  t_{0}^{\ast
}\right)  $ will appear in each Taylor coefficient $\zeta_{m}\left(
t_{0}^{\ast}\right)  $.

\subsection{Structure of the Taylor coefficient $\zeta_{m}\left(  t_{0}^{\ast
}\right)  $}

With $A=\xi_{0}e^{-\theta\;\zeta_{0}\left(  t_{0}^{\ast}\right)  }$,
$Z=1-\zeta_{0}\left(  t_{0}^{\ast}\right)  $ and $x=\theta\,Z$, we list a few
iterations of the recursion (\ref{recursion_zeta_m}),%
\begin{align*}
\zeta_{1}\left(  t_{0}^{\ast}\right)   &  =-A+Z\\
\zeta_{2}\left(  t_{0}^{\ast}\right)   &  =-\frac{A\,^{2}\theta}{2}+\frac
{A}{2!}\left(  x+1\right)  \,-\frac{Z}{2!}\\
\zeta_{3}\left(  t_{0}^{\ast}\right)   &  =-\frac{A^{3}\theta^{2}}{3}%
+\frac{A^{2}\theta}{3!}(3x+2)-\frac{A}{3!}(x+1)^{2}+\frac{Z}{3!}\\
\zeta_{4}\left(  t_{0}^{\ast}\right)   &  =-\frac{A^{4}\theta^{3}}{4}%
+\frac{A^{3}\theta^{2}}{4!}(12x+7)-\frac{A^{2}\theta}{4!}\left(
7x^{2}+11x+3\right)  +\frac{A}{4!}\left(  x^{3}+4x^{2}+3x+1\right)  -\frac
{Z}{4!}\\
\zeta_{5}\left(  t_{0}^{\ast}\right)   &  =-\frac{A^{5}\theta^{4}}{5}%
+\frac{A^{4}\theta^{3}}{5!}(60x+33)-\frac{A^{3}\theta^{2}}{5!}\left(
50x^{2}+69x+17\right) \\
&  +\frac{A^{2}\theta}{5!}\left(  15x^{3}+43x^{2}+28x+4\right)  -\frac{A}%
{5!}\left(  x^{4}+7x^{3}+11x^{2}+4x+1\right)  +\frac{Z}{5!}\\
\zeta_{6}\left(  t_{0}^{\ast}\right)   &  =-\frac{A^{6}\theta^{5}}{6}%
+\frac{A^{5}\theta^{4}}{6!}24(15x+8)-\frac{A^{4}\theta^{3}}{6!}\left(
390x^{2}+499x+120\right) \\
&  +\frac{A^{3}\theta^{2}}{6!}2\left(  90x^{3}+219x^{2}+131x+18\right)
-\frac{A^{2}\theta}{6!}\left(  31x^{4}+142x^{3}+174x^{2}+62x+5\right) \\
&  +\frac{A}{6!}\left(  x^{5}+11x^{4}+32x^{3}+26x^{2}+5x+1\right)  -\frac
{Z}{6!}\\
\zeta_{7}\left(  t_{0}^{\ast}\right)   &  =-\frac{A^{7}\theta^{6}}{7}%
+\frac{A^{6}\theta^{5}}{7!}120(21x+11)-\frac{A^{5}\theta^{4}}{7!}\left(
3360x^{2}+4096x+979\right) \\
&  +\frac{A^{4}\theta^{3}}{7!}\left(  2100x^{3}+4630x^{2}+2641x+370\right)
-\frac{A^{3}\theta^{2}}{7!}2\left(  301x^{4}+1131x^{3}+1218x^{2}%
+421x+36\right) \\
&  +\frac{A^{2}\theta}{7!}\left(  63x^{5}+424x^{4}+850x^{3}+594x^{2}%
+129x+6\right) \\
&  -\frac{A}{7!}\left(  x^{6}+16x^{5}+76x^{4}+122x^{3}+57x^{2}+6x+1\right)
+\frac{Z}{7!}%
\end{align*}
which suggest that%
\begin{equation}
\zeta_{m}\left(  t_{0}^{\ast}\right)  =\frac{\left(  -1\right)  ^{m-1}Z}%
{m!}-\frac{\left(  A\theta\right)  ^{m}}{\theta m}+\frac{\left(  -1\right)
^{m-1}}{\theta m!}\sum_{j=1}^{m-1}\left(  -A\theta\right)  ^{j}p\left(
x;m,j\right)  \label{Taylor_coefficient_zeta_rond_t0}%
\end{equation}
where
\begin{equation}
p\left(  x;m,j\right)  =\sum_{k=0}^{m-j}a_{k}\left(  m,j\right)  x^{k}
\label{def_pol_zeta_m(t_0)}%
\end{equation}
is a polynomial of degree $m-j$ in $x$ with integer coefficients $a_{k}\left(
m,j\right)  $, where $1\leq j\leq m-1$. Around any time point $t_{0}^{\ast}$,
the Taylor coefficient $\zeta_{m}\left(  t_{0}^{\ast}\right)  $ possesses a
general form, where only $A,Z$ and $x$ change with $\zeta_{0}\left(
t_{0}^{\ast}\right)  =\zeta\left(  t_{0}^{\ast}\right)  $. An explicit
solution requires the general form of the coefficients $a_{k}\left(
m,j\right)  $ in the polynomial $p\left(  x;m,j\right)  $, that are
independent of $t_{0}^{\ast}$. The coefficients $a_{k}\left(  m,j\right)  $
are generated by a complicated recursion via (\ref{recursion_zeta_m}) and it
is unlikely that an explicit form can be obtained. For some particular cases,
we give their explicit form in Appendix \ref{sec_coefficients_ak(m,j)}.

\subsection{Taylor series}

Introducing (\ref{Taylor_coefficient_zeta_rond_t0}) in the Taylor series
$\zeta\left(  t^{\ast}\right)  =\zeta_{0}\left(  t_{0}^{\ast}\right)
+\sum_{m=1}^{\infty}\zeta_{m}\left(  t_{0}^{\ast}\right)  \left(  t^{\ast
}-t_{0}^{\ast}\right)  ^{m}$ gives us%
\begin{align*}
\zeta\left(  t^{\ast}\right)   &  =\zeta_{0}\left(  t_{0}^{\ast}\right)
-\frac{1}{\theta}\sum_{m=1}^{\infty}\frac{\left(  A\theta\left(  t^{\ast
}-t_{0}^{\ast}\right)  \right)  ^{m}}{m}-Z\sum_{m=1}^{\infty}\frac{\left(
t_{0}^{\ast}-t\right)  ^{m}}{m!}\\
&  +\frac{1}{\theta}\sum_{m=1}^{\infty}\left(  \sum_{j=1}^{m}\left(
-1\right)  ^{m-1-j}\left(  A\theta\right)  ^{j}p\left(  x;m,j\right)  \right)
\frac{\left(  t^{\ast}-t_{0}^{\ast}\right)  ^{m}}{m!}%
\end{align*}
Provided that $\left\vert A\theta\left(  t^{\ast}-t_{0}^{\ast}\right)
\right\vert <1$, we obtain, with $A=\xi_{0}e^{-\theta\;\zeta_{0}\left(
t_{0}^{\ast}\right)  }$, $Z=1-\zeta_{0}\left(  t_{0}^{\ast}\right)  $ and
$x=\theta Z$, Taylor series of the removed fraction $\zeta\left(  t^{\ast
}\right)  $ around the scaled time $t_{0}^{\ast}$,
\begin{align}
\zeta\left(  t^{\ast}\right)   &  =\zeta_{0}\left(  t_{0}^{\ast}\right)
+Z\left(  1-e^{t_{0}^{\ast}-t^{\ast}}\right)  -\frac{1}{\theta}\log\left(
1-\theta A\left(  t^{\ast}-t_{0}^{\ast}\right)  \right) \nonumber\\
&  -\frac{1}{\theta}\sum_{m=1}^{\infty}\left(  \sum_{j=1}^{m}\left(
-A\theta\right)  ^{j}\sum_{k=0}^{m-j}a_{k}\left(  m,j\right)  x^{k}\right)
\frac{\left(  t_{0}^{\ast}-t^{\ast}\right)  ^{m}}{m!}
\label{Taylor_series_zero_rond_t0}%
\end{align}
Assuming that $p\left(  x;m,j\right)  =O\left(  m^{a}m!\right)  $ for finite
$a$, then the radius $R$ of convergence of the Taylor series $\zeta\left(
t^{\ast}\right)  =\sum_{k=0}^{\infty}\zeta_{k}\left(  t_{0}^{\ast}\right)
\left(  t^{\ast}-t_{0}^{\ast}\right)  ^{k}$ is $\left\vert t^{\ast}%
-t_{0}^{\ast}\right\vert <R=\frac{e^{\theta\;\zeta_{0}\left(  t_{0}^{\ast
}\right)  }}{\xi_{0}\theta}$. The minimum radius of convergence as function of
the normalized effective infection rate $\theta$ occurs at $\theta_{\min
}=\frac{1}{\zeta_{0}\left(  t_{0}^{\ast}\right)  }$. Within the radius of
convergence, the Taylor series (\ref{Taylor_series_zero_rond_t0}) converges as
quickly as a geometric series. The numerical solution of the differential
equation (\ref{Kermack_McKendrick_exact_dvgl_SIR_removed_items}) with
Mathematica is very accurate. The Taylor series in
(\ref{Taylor_series_zero_rond_t0}) attains 6 digits with about 15 terms when
$\left\vert t^{\ast}-t_{0}^{\ast}\right\vert =1$ for $\xi_{0}=0.6$ and
$\theta=2$ at any $\zeta\left(  t_{0}^{\ast}\right)  $.

If $\zeta\left(  t_{0}^{\ast}\right)  $ is known at one time point
$t_{0}^{\ast}$, all values of $\zeta\left(  t^{\ast}\right)  $ can be
obtained, by analytical continuation \cite{Evgrafov_1965,Titchmarshfunctions},
even if the Taylor series (\ref{Taylor_series_zero_rond_t0}) has a finite
radius of convergence. Indeed, starting from $\left(  t_{0}^{\ast}%
,\zeta\left(  t_{0}^{\ast}\right)  \right)  $, the couple $\left(  t_{1}%
^{\ast},\zeta\left(  t_{1}^{\ast}\right)  \right)  $ is found via the Taylor
series sufficiently accurately, which is fed into the new Taylor series around
$t_{1}^{\ast}$ to produce $\left(  t_{2}^{\ast},\zeta\left(  t_{2}^{\ast
}\right)  \right)  $ and so on. The usual starting expansion point
$t_{0}^{\ast}=0$, for which $\zeta_{0}\left(  t_{0}^{\ast}\right)  =0$ and
thus $A=\xi_{0}$ and $Z=1$. If we choose the step small enough\footnote{The
famous Euler transform, which is a special case of an univalent and conformal
M\"{o}bius transform \cite[Vol. 2]{Sansone}, $w=\frac{az+b}{cz+d}$, and whose
summability is treated by Hardy in \cite[chap. VIII]{Hardy_div},
\begin{equation}
f(z)=f_{0}+\sum_{m=1}^{\infty}\left[  \sum_{k=1}^{m}{\binom{m-1}{k-1}}%
\,f_{k}\,q^{m-k}\right]  \;\left(  \frac{z}{1+qz}\right)  ^{m}
\label{Eulertransform}%
\end{equation}
usually extends the convergence range of $z$ compared to the corresponding
Taylor series $f(z)=f_{0}+\sum_{m=1}^{\infty}f_{m}z^{m}$. Here, we set the
Euler transform aside, because numerical computation is not our main aim.},
say $t_{k}^{\ast}-t_{k-1}^{\ast}=\frac{1}{10}$ for $k>1$, then the above
explicitly listed coefficients $\zeta_{m}\left(  t_{0}^{\ast}\right)  $ up to
$O\left(  \left(  t^{\ast}-t_{0}^{\ast}\right)  ^{8}\right)  $ may provide a
sufficient accuracy for each $\zeta\left(  t_{k}^{\ast}\right)  $. The Taylor
series (\ref{scaled_time_converging_series_fractions}) of the inverse function
couples a chosen value of $\zeta$ to the corresponding time $t_{0}^{\ast}$,
whereas the Taylor series $\zeta\left(  t^{\ast}\right)  =\sum_{k=0}^{\infty
}\zeta_{k}\left(  t_{0}^{\ast}\right)  \left(  t^{\ast}-t_{0}^{\ast}\right)
^{k}$ returns $\zeta$ for a chosen value $t^{\ast}$.

\section{Conclusion}

After an overview of the McKendrick differential equations with constant rates
$\beta$ and $\delta$ in (\ref{Kermack_McKendrick_SIR_dvgl}), we have presented
a formal exact solution (at the end of Section
\ref{sec_SIR_governing_equations_solution}) and bounds for the scaled time
$t^{\ast}$ (Section \ref{sec_bounds}). A Taylor series-based approach to
subsequentially approximate the integral
(\ref{Kermack_McKendrick_integral_SIR_removed_fractions}) for the scaled time
$t^{\ast}$ in the SIR epidemic process is presented. The method allows
analytic evaluation up to any desired accuracy, at the expense of many terms.
Similarly, the Taylor series $\zeta\left(  t^{\ast}\right)  =\sum
_{k=0}^{\infty}\zeta_{k}\left(  t_{0}^{\ast}\right)  \left(  t^{\ast}%
-t_{0}^{\ast}\right)  ^{k}$ is derived around $t_{0}^{\ast}$. The
corresponding Taylor coefficients $\zeta_{k}\left(  t_{0}^{\ast}\right)  $ can
be recursively computed up to any order, but the explicit form of $\zeta
_{k}\left(  t_{0}^{\ast}\right)  $ for any $k$ has not been found.

\medskip\textbf{Acknowledgements} I am very grateful to M. Achterberg and B.
Prasse for pointing me to errors.

{\footnotesize
\bibliographystyle{unsrt}
\bibliography{cac,MATH,misc,net,pvm,QTH,tel}
}

\appendix{}

\section{The Lambert function}

\label{sec_Lambert_function}The function $1-\xi_{0}e^{-\theta\zeta}-\zeta$ is
negative if $\zeta>\widetilde{\zeta}$, where $\widetilde{\zeta}$ is the zero
that obeys $1-\xi_{0}e^{-\theta\widetilde{\zeta}}=\widetilde{\zeta}$. With
$u=1-\widetilde{\zeta}$, we rewrite that equation as%
\[
u=1-\widetilde{\zeta}=\xi_{0}e^{-\theta}e^{\theta\left(  1-\widetilde{\zeta
}\right)  }=\xi_{0}e^{-\theta}e^{\theta v}%
\]
or%
\[
\theta ve^{-\theta v}=\theta\xi_{0}e^{-\theta}=a
\]
where $a=\theta\xi_{0}e^{-\theta}\leq\xi_{0}$ is positive real number in
$\left[  0,1\right]  $. Finally, let $q=\theta u=\theta\left(  1-\widetilde
{\zeta}\right)  $, then we arrive at simplest possible form%
\[
qe^{-q}=a
\]
In terms of the Lambert function $v=W\left(  z\right)  $, whose inverse
function is $z=W^{-1}\left(  v\right)  =ve^{v}$, the above equation for the
zero is $W^{-1}\left(  -q\right)  =-a$, which is equivalent to $q=-W\left(
-a\right)  $. Hence, the zero $\widetilde{\zeta}=1-\frac{q}{\theta}$ equals
\begin{equation}
\widetilde{\zeta}=\zeta_{\max}=1+\frac{1}{\theta}W\left(  -\theta\xi
_{0}e^{-\theta}\right)  \label{max_fraction_removed_Lambert}%
\end{equation}
The Lambert function $v=W\left(  z\right)  $, its applications and history is
discussed by Corless \emph{et al.} \cite{Corless1996}. Physically, the zero
$\widetilde{\zeta}$ equals the maximum possible removed fraction $\zeta_{\max
}$ that is reached after infinitely long time when $\frac{dz}{dt^{\ast}}=0$
and the integrand $\frac{1}{1-\xi_{0}e^{-\theta w}-w}$ reaches the real pole
at $w=\widetilde{\zeta}$. If $\theta$ is small, then the zero $\widetilde
{\zeta}=1-\xi_{0}e^{-\theta\widetilde{\zeta}}\simeq1-\xi_{0}$, while if
$\theta$ is large, then $\widetilde{\zeta}\simeq1$. If $\xi_{0}=1$, then
$\widetilde{\zeta}=1-e^{-\theta\widetilde{\zeta}}$, which has the zero
solution $\widetilde{\zeta}=0$, only if $\theta\leq1$. Indeed, the inequality
\cite[p. 103]{PVM_PAComplexNetsCUP}, $e^{-\theta\widetilde{\zeta}}%
<1-\theta\widetilde{\zeta}+\frac{1}{2}\theta^{2}\widetilde{\zeta}^{2}$, leads
to the bound%
\[
\theta\widetilde{\zeta}-\frac{1}{2}\theta^{2}\widetilde{\zeta}^{2}%
<1-e^{-\theta\widetilde{\zeta}}=\widetilde{\zeta}%
\]
which reduces, provided that $\widetilde{\zeta}\neq0$, to the inequality%
\[
\widetilde{\zeta}>\frac{2\left(  \theta-1\right)  }{\theta^{2}}%
\]
that is feasible only if $\theta>1$. If $\theta>1$ and small, then the above
bound is an accurate estimate for $\widetilde{\zeta}$ in
(\ref{max_fraction_removed_Lambert}).

\subsection{Complex zeros of $qe^{-q}=a$ for $a\in\left[  0,1\right]  $}

\label{sec_complex_zeros_qexp(-q)=a}We will determine all complex numbers
$q=x+iy$ that satisfy $qe^{-q}=a$ subject to $0\leq\arg q\leq2\pi$. After
separating real and imaginary part in $\left(  x+iy\right)  =ae^{x+iy}$, we
obtain%
\[
\left\{
\begin{array}
[c]{c}%
x=ae^{x}\cos y\\
y=ae^{x}\sin y
\end{array}
\right.
\]
Their ratio is%
\[
x=y\cot y
\]
and $y=ae^{x}\sin y$ shows that $y=0$ is a solution corresponding to
$x=ae^{x}$. From the last equation, we can eliminate $x=\log\frac{y}{a\sin y}$
and substitute in their ratio,%
\[
\log\frac{y}{a\sin y}=y\cot y
\]
which is even in $y$, but only numerically solvable for $y$.

Further, using $\cos^{2}y+\sin^{2}y=1$ results in a circle around the origin
with radius $ae^{x}$ or $y^{2}=a^{2}e^{2x}-x^{2}=\left(  ae^{x}-x\right)
\left(  ae^{x}+x\right)  $. Since $y$ is real, we either have (a)
$ae^{x}-x\geq0$ and $ae^{x}+x\geq0$ or (b) $ae^{x}-x\leq0$ and $ae^{x}+x\leq
0$. The set (a) is equivalent to $a\geq xe^{-x}$ and $-a\leq xe^{-x}$,
implying that $x>0$ (because $a>0$). The set (b), $0<ae^{x}\leq x$ and
$0>-ae^{x}\geq x$ is not possible. Introducing $y=\pm\sqrt{a^{2}e^{2x}-x^{2}}$
into $x=y\cot y$ yields%
\[
x=\sqrt{a^{2}e^{2x}-x^{2}}\cot\sqrt{a^{2}e^{2x}-x^{2}}%
\]
The plot of the last equation shows that all solutions for $x$ are positive
and the number of solutions grows exponentially fast with $x$! Hence, there
are infinitely many complex zeros. For each positive solution $x$, there are
two values for $y$, symmetric around the real-axis. In other words, the zeros
appear in complex conjugate pairs.

The equations can be expressed in terms of the Lambert function. We rewrite
the first equation as%
\[
-a\cos y=-xe^{-x}=W^{-1}\left(  -x\right)
\]
from which%
\[
x=-W\left(  a\cos y\right)
\]
If $0\leq a\cos y\leq a\leq1$, then $x\in(-W\left(  1\right)  ,0]=(-0.567,0]$.
If $-\frac{1}{e}\leq a\cos y\leq0$, then there are two solutions for $x$,
either $x\in\lbrack0,1]$ or $x>1$. Substituted into $x=y\cot y$, then yields%
\[
y=-W\left(  a\cos y\right)  \tan y
\]
Unfortunately, there is no elegant closed form for a complex zero.

\subsection{The integral
(\ref{Kermack_McKendrick_integral_SIR_removed_fractions}) in terms of the
Lambert function}

Using the derivative of $W\left(  W^{-1}\left(  v\right)  \right)  =v$, we
obtain%
\[
\left.  \frac{dW\left(  x\right)  }{dx}\right\vert _{x=W^{-1}\left(  v\right)
=ve^{v}}=\frac{1}{\frac{dW^{-1}\left(  v\right)  }{dv}}=\frac{1}{e^{v}+ve^{v}%
}=\frac{1}{e^{v}+x}=\frac{1}{\frac{x}{v}+x}%
\]
Thus, with $W\left(  x\right)  =v$ that obeys $x=W\left(  x\right)
e^{W\left(  x\right)  }$, we arrive at
\[
\frac{dW\left(  x\right)  }{dx}=\frac{1}{e^{W\left(  x\right)  }+x}=\frac
{1}{x}\frac{W\left(  x\right)  }{1+W\left(  x\right)  }%
\]
Reconsidering the integral
(\ref{Kermack_McKendrick_integral_SIR_removed_fractions})%
\begin{align*}
t^{\ast}  &  =\int_{0}^{\zeta}\frac{dw}{1-\xi_{0}e^{-\theta w}-w}=\int
_{0}^{\zeta}\frac{dw}{\left(  1-w\right)  -\xi_{0}e^{-\theta}e^{\theta\left(
1-w\right)  }}=\int_{1-\zeta}^{1}\frac{dv}{v-\xi_{0}e^{-\theta}e^{\theta v}}\\
&  =\int_{\theta\left(  1-\zeta\right)  }^{\theta}\frac{du}{u-ae^{u}}%
=-\int_{\theta\left(  1-\zeta\right)  }^{\theta}\frac{e^{-u}du}{a-ue^{-u}%
}=-\int_{\theta\left(  1-\zeta\right)  }^{\theta}\frac{e^{-u}du}%
{a+W^{-1}\left(  -u\right)  }%
\end{align*}
where $a=\xi_{0}\theta e^{-\theta}=-\xi_{0}W^{-1}\left(  -\theta\right)  $.
Let $x=W^{-1}\left(  -u\right)  $, then $u=-W\left(  x\right)  $ and%
\[
t^{\ast}=\int_{W^{-1}\left(  -\theta\left(  1-\zeta\right)  \right)  }%
^{W^{-1}\left(  -\theta\right)  }\frac{e^{W\left(  x\right)  }\frac{dW\left(
x\right)  }{dx}dx}{a+x}=\int_{W^{-1}\left(  -\theta\left(  1-\zeta\right)
\right)  }^{W^{-1}\left(  -\theta\right)  }\frac{dx}{\left(  1+xe^{-W\left(
x\right)  }\right)  \left(  a+x\right)  }%
\]
Finally, with $x=W\left(  x\right)  e^{W\left(  x\right)  }$ and
$W^{-1}\left(  x\right)  =xe^{x}$, we arrive at%
\[
t^{\ast}=\int_{\theta e^{-\theta}}^{\theta\left(  1-\zeta\right)
e^{-\theta\left(  1-\zeta\right)  }}\frac{dx}{\left(  1+W\left(  -x\right)
\right)  \left(  \xi_{0}\theta e^{-\theta}-x\right)  }%
\]

We mention another possible route. Since $\frac{d}{du}\left(  ue^{-u}%
-a\right)  =-ue^{-u}+e^{-u}=\left(  1-u\right)  e^{-u}$, we have%
\[
t^{\ast}=\int_{\theta\left(  1-\zeta\right)  }^{\theta}\frac{e^{-u}du}%
{ue^{-u}-a}=\int_{\theta\left(  1-\zeta\right)  }^{\theta}\frac{d\left(
ue^{-u}-a\right)  }{\left(  1-u\right)  \left(  ue^{-u}-a\right)  }%
\]
Partial integration yields%
\[
t^{\ast}=\frac{\log\left(  \theta e^{-\theta}\left(  1-\xi_{0}\right)
\right)  }{1-\theta}-\frac{\log\left(  \theta e^{-\theta}\left(  \left(
1-\zeta\right)  e^{\theta\zeta}-\xi_{0}\right)  \right)  }{1-\theta\left(
1-\zeta\right)  }-\int_{\theta\left(  1-\zeta\right)  }^{\theta}\frac
{\log\left(  ue^{-u}-a\right)  }{\left(  1-u\right)  ^{2}}du
\]
which can be problematic if $\theta\left(  1-\zeta\right)  <1$ and $\theta>1$,
due to the pole at $u=1$. Invoking contour integration -- Cauchy's principal
value -- can be considered.

\section{Characteristic coefficients of a complex function}

\label{sec_chc}If $f\left(  z\right)  $ has a Taylor series around $z_{0}$,%
\[
f\left(  z\right)  =\sum_{k=0}^{\infty}f_{k}\left(  z_{0}\right)  \left(
z-z_{0}\right)  ^{k}\hspace{1cm}\text{with }f_{k}\left(  z_{0}\right)
=\frac{1}{k!}\left.  \frac{d^{k}f\left(  z\right)  }{dz^{k}}\right\vert
_{z=z_{0}}%
\]
then the general relation where $G\left(  z\right)  $ is analytic around
$f\left(  z_{0}\right)  $ is%
\begin{equation}
G(f(z))=G(f\left(  z_{0}\right)  )+\sum_{m=1}^{\infty}\left(  \sum_{k=1}%
^{m}\frac{1}{k!}\;\left.  \frac{d^{k}G(p)}{dp^{k}}\right\vert _{p=f(z_{0}%
)}\,s[k,m]_{f(z)}(z_{0})\right)  \left(  z-z_{0}\right)  ^{m}
\label{chc_G(f(z))_rond_z0}%
\end{equation}
where the characteristic coefficient \cite{PVM_ASYM} of a complex function
$f\left(  z\right)  $ has the combinatorial form
\[
s[k,m]_{f(z)}\left(  z_{0}\right)  =\sum_{\sum_{i=1}^{k}j_{i}=m;j_{i}>0}%
\prod_{i=1}^{k}f_{j_{i}}(z_{0})
\]
which obeys the recursion relation%
\begin{align}
s[1,m]_{f(z)}\left(  z_{0}\right)   &  =f_{m}\left(  z_{0}\right) \nonumber\\
s[k,m]_{f(z)}\left(  z_{0}\right)   &  =\sum_{j=1}^{m-k+1}f_{j}\;\left(
z_{0}\right)  s[k-1,m-j]_{f(z)}\left(  z_{0}\right)  \hspace{2cm}(k>1)
\label{s_recursive}%
\end{align}
For $k\leq m$ and $m>0$, the characteristic coefficient of a function $f(z)$
around $z_{0}$ also equals
\begin{equation}
s[k,m]_{f(z)}\left(  z_{0}\right)  =\frac{1}{m!}\,\left.  \frac{d^{m}}{dz^{m}%
}[f(z)-f(z_{0})]^{k}\right\vert _{z=z_{0}} \label{s_general}%
\end{equation}
illustrating that $s[k,m]_{f(z)}\left(  z_{0}\right)  =0$ for a constant
function. The characteristic coefficient $s[k,m]_{f(z)}\left(  z_{0}\right)  $
is a fundamental building block in the theory of generalized Taylor series.
Clearly, (\ref{chc_G(f(z))_rond_z0}) reduces to Taylor series of $G\left(
z\right)  $ for $f\left(  z\right)  =z$ and, thus, $s[k,m]_{z}\left(
z_{0}\right)  =\delta_{k,m}$.

\section{Proof of Theorem \ref{theorem_Taylor_series_SIR_time}}

\label{sec_proof_Taylor_expansion_time_in_removed}We present three proofs, a
direct computation involving our characteristic coefficients (Section
\ref{sec_chc}), a verification proof, that avoids characteristic coefficients
and a proof based on repeated partial integrations.

A) If the Taylor series of a complex function $f\left(  z\right)  =\sum
_{k=0}^{\infty}f_{k}\left(  z_{0}\right)  \,\left(  z-z_{k}\right)  ^{k}$,
then%
\begin{equation}
\frac{1}{f(z)}=\frac{1}{f_{0}\left(  z_{0}\right)  }+\sum_{m=1}^{\infty
}\left[  \sum_{k=1}^{m}\frac{(-1)^{k}}{\left(  f_{0}\left(  z_{0}\right)
\right)  ^{k+1}}\,s[k,m]\left(  z_{0}\right)  \right]  \,\left(
z-z_{0}\right)  ^{m} \label{1opf}%
\end{equation}
where $s\left[  k,m\right]  \left(  z_{0}\right)  $ is the characteristic
coefficient of the function $f\left(  z\right)  $ around $z_{0}$.

The Taylor series of the entire function $h\left(  z\right)  =1-\xi
_{0}e^{-\theta z}-z$ of the complex variable $z$ around $z_{0}$ is%
\begin{align*}
h\left(  z\right)   &  =1-\xi_{0}e^{-\theta z}-z=1-z_{0}-\left(
z-z_{0}\right)  -\xi_{0}e^{-\theta z_{0}}e^{-\theta\left(  z-z_{0}\right)  }\\
&  =1-\xi_{0}e^{-\theta z_{0}}-z_{0}+\left(  \theta\xi_{0}e^{-\theta z_{0}%
}-1\right)  \left(  z-z_{0}\right)  -\xi_{0}e^{-\theta z_{0}}\sum
_{k=2}^{\infty}\frac{\left(  -\theta\right)  ^{k}}{k!}\left(  z-z_{0}\right)
^{k}%
\end{align*}
Since the characteristic coefficient of $e^{z}$ around $z_{0}=0$ is known as
\begin{equation}
\mathcal{S}_{m}^{(k)}=\frac{m!}{k!}\left.  s[k,m]\right\vert _{e^{z}}\left(
0\right)  \label{StirlingS2_chc_exp(z)}%
\end{equation}
where $\mathcal{S}_{m}^{(k)}$ is the Stirling numbers of the second kind
\cite{Abramowitz}, we apply the property
\begin{equation}
\left.  s[k,m]\right\vert _{f(\alpha\,z)}=\alpha^{m}\,\left.
s[k,m]\right\vert _{f(z)} \label{scaling_argument}%
\end{equation}
to obtain%
\[
\left.  s[k,m]\right\vert _{e^{-\tau z}}=\left(  -\tau\right)  ^{m}\frac
{k!}{m!}\mathcal{S}_{m}^{(k)}%
\]
From (\ref{s_general}), it follows that $\left.  s[k,m]\right\vert _{e^{-\tau
z}}\left(  z_{0}\right)  =e^{-k\tau z_{0}}\left.  s[k,m]\right\vert _{e^{-\tau
z}}\left(  0\right)  $ and%
\[
\left.  s[k,m]\right\vert _{e^{-\tau z}}\left(  z_{0}\right)  =e^{-k\tau
z_{0}}\left(  -\tau\right)  ^{m}\frac{k!}{m!}\mathcal{S}_{m}^{(k)}%
\]
Next, the characteristic coefficient of $N-z$ follows directly from
(\ref{s_general})%
\[
\left.  s[k,m]\right\vert _{N-z}\left(  z_{0}\right)  =\left(  -1\right)
^{k}1_{\left\{  k=m\right\}  }%
\]
With a little more effort, we find that%
\begin{equation}
\left.  s[k,m]\right\vert _{\alpha f(z)+\beta g(z)}(z_{0})=\,\sum_{j=0}%
^{k}\binom{k}{j}\alpha^{k-j}\beta^{j}\sum_{n=0}^{m}\left.
s[k-j,m-n]\right\vert _{f\left(  z\right)  }(z_{0})\left.  s[j,n]\right\vert
_{g\left(  z\right)  }(z_{0}) \label{chc_sum_korter}%
\end{equation}
Applying (\ref{chc_sum_korter}) to $g\left(  z\right)  =1-z$ and
$f(z)=e^{-\theta z}$ with $\alpha=\left(  -\xi_{0}\right)  $ yields
\begin{align*}
\left.  s[k,m]\right\vert _{1-\zeta-\xi_{0}e^{-\theta\zeta}}\left(
z_{0}\right)   &  =\,\sum_{j=0}^{k}\binom{k}{j}\left(  -\xi_{0}\right)
^{k-j}\sum_{n=0}^{m}\left.  s[k-j,m-n]\right\vert _{e^{-\theta z}}\left.
s[j,n]\right\vert _{1-z}\\
&  =\left(  -\xi_{0}\right)  ^{k}\sum_{j=0}^{k}\binom{k}{j}\left(  -\xi
_{0}\right)  ^{-j}\sum_{n=0}^{m}e^{-\left(  k-j\right)  \theta z_{0}}\left(
-\theta\right)  ^{m-n}\frac{\left(  k-j\right)  !}{\left(  m-n\right)
!}\mathcal{S}_{m-n}^{(k-j)}\left(  -1\right)  ^{j}1_{\left\{  j=n\right\}  }%
\end{align*}
and%
\begin{equation}
\left.  s[k,m]\right\vert _{1-\zeta-\xi_{0}e^{-\theta\zeta}}\left(
z_{0}\right)  =\frac{k!}{m!}\left(  -1\right)  ^{k}\sum_{j=0}^{k}\binom{m}%
{j}\left(  \xi_{0}e^{-\theta z_{0}}\right)  ^{k-j}\left(  -\theta\right)
^{m-j}\mathcal{S}_{m-j}^{(k-j)} \label{chc_h_rond_z0}%
\end{equation}

We are now ready to apply (\ref{1opf})%
\begin{equation}
\frac{1}{1-\zeta-\xi_{0}e^{-\theta\zeta}}=\frac{1}{1-z_{0}-\xi_{0}e^{-\theta
z_{0}}}+\sum_{m=1}^{\infty}\left[  \sum_{k=1}^{m}\frac{k!\sum_{j=0}^{k}%
\binom{m}{j}\left(  \xi_{0}e^{-\theta z_{0}}\right)  ^{k-j}\left(
-\theta\right)  ^{m-j}\mathcal{S}_{m-j}^{(k-j)}}{\left(  1-z_{0}-\xi
_{0}e^{-\theta z_{0}}\right)  ^{k+1}}\right]  \,\frac{\left(  \zeta
-z_{0}\right)  ^{m}}{m!} \label{Taylor_1_op_h(z)}%
\end{equation}
Finally, $t^{\ast}=H\left(  \zeta\right)  =\int_{0}^{z}\frac{du}{1-u-\xi
_{0}e^{-\theta u}}$ follows after integration of the Taylor series
(\ref{Taylor_1_op_h(z)}) as%
\begin{equation}
t^{\ast}=H\left(  \zeta\right)  =\frac{\zeta}{1-z_{0}-\xi_{0}e^{-\theta z_{0}%
}}+\sum_{m=1}^{\infty}\left[  \sum_{k=1}^{m}\frac{k!\sum_{j=0}^{k}\binom{m}%
{j}\left(  \xi_{0}e^{-\theta z_{0}}\right)  ^{k-j}\left(  -\theta\right)
^{m-j}\mathcal{S}_{m-j}^{(k-j)}}{\left(  1-z_{0}-\xi_{0}e^{-\theta z_{0}%
}\right)  ^{k+1}}\right]  \,\frac{\left(  \zeta-z_{0}\right)  ^{m+1}-\left(
-z_{0}\right)  ^{m+1}}{\left(  m+1\right)  !}
\label{Taylor_series_H(z)_rond_z0}%
\end{equation}

The Taylor series (\ref{Taylor_series_H(z)_rond_z0}) converges reasonably fast
if we choose $z_{0}=\frac{\zeta}{2}$, which minimizes both $\left(
\zeta-z_{0}\right)  ^{m+1}$ and $\left(  -z_{0}\right)  ^{m+1}$. In that case,%
\[
\left(  \zeta-z_{0}\right)  ^{m+1}-\left(  -z_{0}\right)  ^{m+1}=\left(
\frac{\zeta}{2}\right)  ^{m+1}-\left(  -\frac{\zeta}{2}\right)  ^{m+1}=\left(
\frac{\zeta}{2}\right)  ^{m+1}\left(  1+\left(  -1\right)  ^{m}\right)
\]
and only even terms in $m$ remain. With the choice $z_{0}=\frac{\zeta}{2}$,
the Taylor series (\ref{Taylor_series_H(z)_rond_z0}) becomes
(\ref{scaled_time_converging_series_fractions}).

B) Reversing the $m$- and $k$-sum in (\ref{Taylor_1_op_h(z)}) gives us%
\begin{align*}
\frac{1}{1-\zeta-\xi_{0}e^{-\theta\zeta}}  &  =\frac{1}{1-z_{0}-\xi
_{0}e^{-\theta z_{0}}}+\sum_{k=1}^{\infty}\frac{k!\sum_{j=0}^{k}\left(
\xi_{0}e^{-\theta z_{0}}\right)  ^{k-j}}{\left(  1-z_{0}-\xi_{0}e^{-\theta
z_{0}}\right)  ^{k+1}}\sum_{m=k}^{\infty}\binom{m}{j}\left(  -\theta\right)
^{m-j}\mathcal{S}_{m-j}^{(k-j)}\,\frac{\left(  \zeta-z_{0}\right)  ^{m}}{m!}\\
&  =\frac{1}{1-z_{0}-\xi_{0}e^{-\theta z_{0}}}+\sum_{k=1}^{\infty}\frac
{k!\sum_{j=0}^{k}\frac{\left(  \xi_{0}e^{-\theta z_{0}}\right)  ^{k-j}}%
{j!}\left(  -\theta\right)  ^{-j}}{\left(  1-z_{0}-\xi_{0}e^{-\theta z_{0}%
}\right)  ^{k+1}}\sum_{m=k}^{\infty}\mathcal{S}_{m-j}^{(k-j)}\,\frac{\left(
\theta\left(  z_{0}-\zeta\right)  \right)  ^{m}}{\left(  m-j\right)  !}%
\end{align*}
Further,%
\[
\sum_{m=k}^{\infty}\mathcal{S}_{m-j}^{(k-j)}\,\frac{\left(  \theta\left(
z_{0}-\zeta\right)  \right)  ^{m}}{\left(  m-j\right)  !}=\left(
\theta\left(  z_{0}-\zeta\right)  \right)  ^{j}\sum_{m=k-j}^{\infty
}\mathcal{S}_{m}^{(k-j)}\,\frac{\left(  \theta\left(  z_{0}-\zeta\right)
\right)  ^{m}}{m!}%
\]
and invoking the generating function of the Stirling Numbers of the Second
Kind \cite[Sec. 24.1.4]{Abramowitz}%
\[
(e^{x}-1)^{k}=k!\,\sum_{m=k}^{\infty}\mathcal{S}_{m}^{(k)}\,\frac{x^{m}}{m!}%
\]
yields%
\[
\sum_{m=k}^{\infty}\mathcal{S}_{m-j}^{(k-j)}\,\frac{\left(  \theta\left(
z_{0}-\zeta\right)  \right)  ^{m}}{\left(  m-j\right)  !}=\left(
\theta\left(  z_{0}-\zeta\right)  \right)  ^{j}\frac{(e^{\theta\left(
z_{0}-\zeta\right)  }-1)^{k-j}}{\left(  k-j\right)  !}%
\]
Hence,%
\begin{align*}
\frac{1}{1-\zeta-\xi_{0}e^{-\theta\zeta}}  &  =\frac{1}{1-z_{0}-\xi
_{0}e^{-\theta z_{0}}}+\sum_{k=1}^{\infty}\frac{\sum_{j=0}^{k}\binom{k}%
{j}\left(  \xi_{0}e^{-\theta z_{0}}(e^{\theta\left(  z_{0}-\zeta\right)
}-1)\right)  ^{k-j}\left(  \zeta-z_{0}\right)  ^{j}}{\left(  1-z_{0}-\xi
_{0}e^{-\theta z_{0}}\right)  ^{k+1}}\\
&  =\frac{1}{1-z_{0}-\xi_{0}e^{-\theta z_{0}}}+\sum_{k=1}^{\infty}%
\frac{\left(  \zeta-z_{0}+\xi_{0}(e^{-\theta\zeta}-e^{-\theta z_{0}})\right)
^{k}}{\left(  1-z_{0}-\xi_{0}e^{-\theta z_{0}}\right)  ^{k+1}}\\
&  =\frac{1}{1-z_{0}-\xi_{0}e^{-\theta z_{0}}}\left(  1+\sum_{k=1}^{\infty
}\left(  \frac{\zeta-z_{0}+\xi_{0}(e^{-\theta\zeta}-e^{-\theta z_{0}}%
)}{1-z_{0}-\xi_{0}e^{-\theta z_{0}}}\right)  ^{k}\right)
\end{align*}
and%
\begin{align*}
\frac{1}{1-\zeta-\xi_{0}e^{-\theta\zeta}}  &  =\frac{1}{1-z_{0}-\xi
_{0}e^{-\theta z_{0}}}\sum_{k=0}^{\infty}\left(  \frac{\left(  1-z_{0}-\xi
_{0}e^{-\theta z_{0}}\right)  -\left(  1-\zeta-\xi_{0}e^{-\theta\zeta}\right)
}{1-z_{0}-\xi_{0}e^{-\theta z_{0}}}\right)  ^{k}\\
&  =\frac{1}{1-z_{0}-\xi_{0}e^{-\theta z_{0}}}\sum_{k=0}^{\infty}\left(
1-\frac{1-\zeta-\xi_{0}e^{-\theta\zeta}}{1-z_{0}-\xi_{0}e^{-\theta z_{0}}%
}\right)  ^{k}\\
&  =\frac{1}{1-z_{0}-\xi_{0}e^{-\theta z_{0}}}\frac{1}{1-1+\frac{1-\zeta
-\xi_{0}e^{-\theta\zeta}}{1-z_{0}-\xi_{0}e^{-\theta z_{0}}}}=\frac{1}%
{1-z_{0}-\xi_{0}e^{-\theta z_{0}}}\frac{1}{\frac{1-\zeta-\xi_{0}%
e^{-\theta\zeta}}{1-z_{0}-\xi_{0}e^{-\theta z_{0}}}}%
\end{align*}
resulting in an identity and demonstrating that the Taylor series
(\ref{Taylor_1_op_h(z)}) is correct. Moreover, convergence requires that
$\left\vert 1-\frac{1-\zeta-\xi_{0}e^{-\theta\zeta}}{1-z_{0}-\xi_{0}e^{-\theta
z_{0}}}\right\vert <1$, which is equivalent in terms of fractions to%
\[
\frac{\zeta-z_{0}+\xi_{0}(e^{-\theta\zeta}-e^{-\theta z_{0}})}{1-z_{0}-\xi
_{0}e^{-\theta z_{0}}}\leq1
\]
or%
\[
\zeta\leq1-\xi_{0}e^{-\theta\zeta}%
\]
which is always satisfied for any (physical) fraction of removed items
$\zeta\leq\zeta_{\max}$, because, as shown in Section
\ref{sec_Lambert_function}, the maximum possible fraction of removed items
$\zeta_{\max}$ satisfies
\[
\zeta_{\max}=1-\xi_{0}e^{-\theta\zeta_{\max}}%
\]
Consequently, all terms in $\sum_{k=0}^{\infty}\left(  1-\frac{1-\zeta-\xi
_{0}e^{-\theta\zeta}}{1-z_{0}-\xi_{0}e^{-\theta z_{0}}}\right)  ^{k}$ are
positive, as well as in the integrated power series.

C) From the general relation \cite{PVM_charcoef}%
\begin{equation}
\int_{a}^{b}f\left(  x\right)  g\left(  x\right)  dx=\int_{a}^{b}\left(
\sum_{k=0}^{m-1}\frac{g^{\left(  k\right)  }\left(  b\right)  }{k!}\left(
u-b\right)  ^{k}\right)  f\left(  u\right)  du+\frac{\left(  -1\right)  ^{m}%
}{\left(  m-1\right)  !}\int_{a}^{b}dx\;g^{\left(  m\right)  }\left(
x\right)  \int_{a}^{x}\left(  x-u\right)  ^{m-1}f\left(  u\right)  du
\label{repeated_partial_integration_single_integral}%
\end{equation}
we find, for $f\left(  u\right)  =1$ and $g\left(  u\right)  =\frac{1}%
{1-u-\xi_{0}e^{-\theta u}}$,%
\begin{align*}
\int_{0}^{\zeta}\frac{du}{1-u-\xi_{0}e^{-\theta u}}  &  =\int_{0}^{\zeta
}\left(  \sum_{k=0}^{m-1}\frac{g^{\left(  k\right)  }\left(  \zeta\right)
}{k!}\left(  u-\zeta\right)  ^{k}\right)  du+\frac{\left(  -1\right)  ^{m}%
}{\left(  m-1\right)  !}\int_{0}^{\zeta}dx\;g^{\left(  m\right)  }\left(
x\right)  \int_{0}^{x}\left(  x-u\right)  ^{m-1}du\\
&  =\sum_{k=0}^{m-1}\frac{g^{\left(  k\right)  }\left(  \zeta\right)  }%
{k!}\int_{0}^{\zeta}\left(  u-\zeta\right)  ^{k}du+\frac{\left(  -1\right)
^{m}}{\left(  m-1\right)  !}\int_{0}^{\zeta}dx\;g^{\left(  m\right)  }\left(
x\right)  \int_{0}^{x}\left(  x-u\right)  ^{m-1}du
\end{align*}
and, with $\int_{0}^{z}\left(  u-z\right)  ^{k}du=\left.  \frac{\left(
u-z\right)  ^{k+1}}{k+1}\right\vert _{0}^{z}=\frac{\left(  -1\right)
^{k}z^{k+1}}{k+1}$,%
\[
\int_{0}^{\zeta}\frac{du}{1-u-\xi_{0}e^{-\theta u}}=\sum_{k=0}^{m-1}%
\frac{\left(  -1\right)  ^{k}g^{\left(  k\right)  }\left(  \zeta\right)
}{\left(  k+1\right)  !}\zeta^{k+1}-\frac{\left(  -1\right)  ^{m}}{m!}\int
_{0}^{\zeta}\;x^{m}g^{\left(  m\right)  }\left(  x\right)  dx
\]
Hence,%
\begin{align*}
\int_{0}^{\zeta}\frac{du}{1-u-\xi_{0}e^{-\theta u}}  &  =\sum_{k=0}%
^{m-1}\left.  \frac{d^{k}}{du^{k}}\left(  \frac{1}{1-u-\xi_{0}e^{-\theta u}%
}\right)  \right\vert _{u=\zeta}\frac{\left(  -1\right)  ^{k}\zeta^{k+1}%
}{\left(  k+1\right)  !}\\
&  -\frac{\left(  -1\right)  ^{m}}{m!}\int_{0}^{\zeta}\;t^{m}\left.
\frac{d^{m}}{du^{m}}\left(  \frac{1}{1-u-\xi_{0}e^{-\theta u}}\right)
\right\vert _{u=t}dt
\end{align*}
which leads for $m\rightarrow\infty$ to the Taylor series
(\ref{Taylor_series_H(z)_rond_z0}) for $z_{0}=\zeta$. Consequently,%
\begin{equation}
\left.  \frac{d^{m}}{du^{m}}\left(  \frac{1}{1-u-\xi_{0}e^{-\theta u}}\right)
\right\vert _{u=\zeta}=\sum_{k=1}^{m}\frac{k!\sum_{j=0}^{k}\binom{m}{j}\left(
\xi_{0}e^{-\theta\zeta}\right)  ^{k-j}\left(  -\theta\right)  ^{m-j}%
\mathcal{S}_{m-j}^{(k-j)}}{\left(  1-\zeta-\xi_{0}e^{-\theta\zeta}\right)
^{k+1}} \label{D^m(1op(N-x-z))}%
\end{equation}
and, also with the characteristic coefficient (\ref{chc_h_rond_z0}),%
\[
\left.  \frac{d^{m}}{du^{m}}\left(  \frac{1}{1-u-\xi_{0}e^{-\theta u}}\right)
\right\vert _{u=\zeta}=m!\sum_{k=1}^{m}\frac{\left(  -1\right)  ^{k}\left.
s[k,m]\right\vert _{1-\zeta-\xi_{0}e^{-\theta\zeta}}\left(  \zeta\right)
}{\left(  1-\zeta-\xi_{0}e^{-\theta\zeta}\right)  ^{k+1}}%
\]

\section{Further developments of the Taylor series}

\label{sec_summing_Taylor_series}

\subsection{Other expression for the characteristic coefficient $\left.
s[k,m]\right\vert _{1-\zeta-\xi_{0}e^{-\theta\zeta}}\left(  z_{0}\right)  $}

Denoting $s^{\ast}[k,m]=\left.  s[k,m]\right\vert _{\forall j\;:\;f_{j}%
\rightarrow f_{j+1}}$, which means that we shift each Taylor coefficients one
upwards, then we can show \cite{PVM_charcoef}, for $m>k$, that
\[
s[k,m]\left(  z_{0}\right)  =\sum_{j=1}^{m-k}{\binom{k}{j}}\;f_{1}%
^{k-j}\,s^{\ast}[j,m-k]\left(  z_{0}\right)
\]
and, in general, $s[m,m]\left(  z_{0}\right)  =f_{1}^{m}\left(  z_{0}\right)
$. For the function $h\left(  z\right)  =1-\xi_{0}e^{-\theta z}-z$, with
Taylor coefficients $h_{0}\left(  z_{0}\right)  =1-z_{0}-\zeta_{0}e^{-\theta
z_{0}}$, $h_{1}\left(  z_{0}\right)  =\xi_{0}\theta e^{-\theta z_{0}}-1$ and
$h_{j}\left(  z_{0}\right)  =-\xi_{0}e^{-\theta z_{0}}\frac{\left(
-\theta\right)  ^{j}}{j!}$ for $j>1$, we have%
\[
\left.  s[k,m]\right\vert _{1-\zeta-\xi_{0}e^{-\theta\zeta}}\left(
z_{0}\right)  =\sum_{j=1}^{m-k}{\binom{k}{j}}\left(  \theta\xi_{0}e^{-\theta
z_{0}}-1\right)  ^{k-j}\,\left.  s^{\ast}[j,m-k]\left(  z_{0}\right)
\right\vert _{1-\zeta-\xi_{0}e^{-\theta\zeta}}%
\]
where $\left.  s^{\ast}[k,m]\left(  z_{0}\right)  \right\vert _{1-\zeta
-\xi_{0}e^{-\theta\zeta}}=\left.  s^{\ast}[k,m]\left(  z_{0}\right)
\right\vert _{-\xi_{0}e^{-\theta\zeta}}$ and $\left.  s[k,m]\right\vert
_{\left(  -\xi_{0}\right)  e^{\theta z}}\left(  z_{0}\right)  =\left(
-1\right)  ^{k+m}\xi_{0}^{k}\theta^{m}e^{k\theta z_{0}}\frac{k!}%
{m!}\mathcal{S}_{m}^{(k)}$. With%
\begin{align*}
\left.  s^{\ast}[k,m]\right\vert _{\left(  -\xi_{0}\right)  e^{\theta z}%
}\left(  z_{0}\right)   &  =\sum_{j=0}^{k}\binom{k}{j}\left(  -1\right)
^{k-j}\left(  \xi_{0}\theta\right)  ^{k-j}e^{-\left(  k-j\right)  \theta
z_{0}}\left.  s[j,m+j]\right\vert _{\left(  -\xi_{0}\right)  e^{\theta z}%
}\left(  z_{0}\right) \\
&  =\left(  -1\right)  ^{m}\sum_{j=0}^{k}\binom{k}{j}\left(  -1\right)
^{k-j}\left(  \xi_{0}\theta\right)  ^{k-j}e^{\left(  k-j\right)  \theta z_{0}%
}\xi_{0}^{j}\theta^{m+j}e^{j\theta z_{0}}\frac{j!}{\left(  m+j\right)
!}\mathcal{S}_{m+j}^{(j)}\\
&  =\left(  -1\right)  ^{m}\xi_{0}^{k}\theta^{m+k}e^{-k\theta z_{0}}%
k!\sum_{j=0}^{k}\frac{\left(  -1\right)  ^{k-j}}{\left(  k-j\right)  !\left(
m+j\right)  !}\mathcal{S}_{m+j}^{(j)}%
\end{align*}
leads, for $k<m$, to%
\[
\left.  s[k,m]\right\vert _{1-\zeta-\xi_{0}e^{-\theta\zeta}}\left(
z_{0}\right)  =\left(  -1\right)  ^{m-k}\theta^{m-k}\sum_{j=1}^{m-k}{\binom
{k}{j}}\left(  \theta\xi_{0}e^{-\theta z_{0}}-1\right)  ^{k-j}\,\left(
\xi_{0}\theta\right)  ^{j}e^{-j\theta z_{0}}j!T\left(  j,m-k\right)
\]
where the sum%
\begin{equation}
T\left(  j,m\right)  =\sum_{q=0}^{j}\frac{\left(  -1\right)  ^{j-q}}{\left(
j-q\right)  !\left(  m+q\right)  !}\mathcal{S}_{m+q}^{(q)}=\frac{\left(
-1\right)  ^{j}}{j!}\delta_{m0}+\sum_{q=1}^{j}\frac{\left(  -1\right)  ^{j-q}%
}{\left(  j-q\right)  !\left(  m+q\right)  !}\mathcal{S}_{m+q}^{(q)}
\label{def_sum_T(j,m)}%
\end{equation}
is always positive and equals the $\left.  s[k,m]\right\vert _{ze^{-\tau z}%
}\left(  0\right)  $. From (\ref{chc_h_rond_z0}), we find%
\[
\left.  s[m,m]\right\vert _{1-\zeta-\xi_{0}e^{-\theta\zeta}}\left(
z_{0}\right)  =\left(  -1\right)  ^{m}\sum_{j=0}^{m}\binom{m}{j}\left(
-\theta\xi_{0}e^{-\theta z_{0}}\right)  ^{m-j}=\left(  \theta\xi_{0}e^{-\theta
z_{0}}-1\right)  ^{m}%
\]

We are now ready to apply (\ref{1opf})%
\begin{align}
\frac{1}{1-\zeta-\xi_{0}e^{-\theta\zeta}}  &  =\frac{1}{1-z_{0}-\xi
_{0}e^{-\theta z_{0}}}\label{Taylor_series_second_chc}\\
&  +\sum_{m=1}^{\infty}\left[  \sum_{k=1}^{m}\frac{\theta^{m-k}\sum
_{j=1}^{m-k}{\binom{k}{j}}\left(  \theta\xi_{0}e^{-\theta z_{0}}-1\right)
^{k-j}\,\left(  \xi_{0}\theta\right)  ^{j}e^{-j\theta z_{0}}j!T\left(
j,m-k\right)  }{\left(  1-z_{0}-\xi_{0}e^{-\theta z_{0}}\right)  ^{k+1}%
}\right]  \,\left(  z_{0}-\zeta\right)  ^{m}\nonumber
\end{align}

\subsection{Splitting off the $k=m$ term in (\ref{Taylor_series_H(z)_rond_z0}%
)}

We split-off the $k=m$ term in the Taylor series
(\ref{Taylor_series_H(z)_rond_z0}) around the point $z_{0}$,%
\begin{align*}
t^{\ast}  &  =\frac{\zeta}{1-z_{0}-\xi_{0}e^{-\theta\zeta_{0}}}+\sum
_{m=1}^{\infty}\frac{\,\left(  1-\theta\xi_{0}e^{-\theta z_{0}}\right)  ^{m}%
}{\left(  1-z_{0}-\xi_{0}e^{-\theta z_{0}}\right)  ^{m+1}}\frac{\left(
\zeta-z_{0}\right)  ^{m+1}}{\left(  m+1\right)  }\\
&  -\sum_{m=1}^{\infty}\frac{\,\left(  1-\theta\xi_{0}e^{-\theta z_{0}%
}\right)  ^{m}}{\left(  1-z_{0}-\xi_{0}e^{-\theta z_{0}}\right)  ^{m+1}}%
\frac{\left(  -z_{0}\right)  ^{m+1}}{\left(  m+1\right)  }\\
&  +\sum_{m=1}^{\infty}\left[  \sum_{k=1}^{m-1}\frac{k!\,\sum_{j=0}^{k}%
\binom{m}{j}\left(  \xi_{0}e^{-\theta z_{0}}\right)  ^{k-j}\left(
-\theta\right)  ^{m-j}\mathcal{S}_{m-j}^{(k-j)}}{\left(  1-z_{0}-\xi
_{0}e^{-\theta z_{0}}\right)  ^{k+1}}\right]  \frac{\left(  \zeta
-z_{0}\right)  ^{m+1}-\left(  -z_{0}\right)  ^{m+1}}{\left(  m+1\right)  !}%
\end{align*}
It holds that%
\begin{align*}
Y\left(  y\right)   &  =\sum_{m=1}^{\infty}\frac{\,\left(  1-\theta\xi
_{0}e^{-\theta z_{0}}\right)  ^{m}}{\left(  1-z_{0}-\xi_{0}e^{-\theta z_{0}%
}\right)  ^{m+1}}\frac{y^{m+1}}{\left(  m+1\right)  }=\sum_{m=2}^{\infty}%
\frac{\,\left(  1-\theta\xi_{0}e^{-\theta z_{0}}\right)  ^{m-1}}{\left(
1-z_{0}-\xi_{0}e^{-\theta z_{0}}\right)  ^{m}}\frac{y^{m}}{m}\\
&  =\frac{1}{1-\theta\xi_{0}e^{-\theta z_{0}}}\sum_{m=2}^{\infty}\frac{1}%
{m}\left(  \frac{\,y\left(  1-\theta\xi_{0}e^{-\theta z_{0}}\right)  }%
{1-z_{0}-\xi_{0}e^{-\theta z_{0}}}\right)  ^{m}\\
&  =\frac{1}{1-\theta\xi_{0}e^{-\theta z_{0}}}\left\{  -\ln\left(
1-\frac{\,y\left(  1-\theta\xi_{0}e^{-\theta z_{0}}\right)  }{1-z_{0}-\xi
_{0}e^{-\theta z_{0}}}\right)  -\frac{\,y\left(  1-\theta\xi_{0}e^{-\theta
z_{0}}\right)  }{1-z_{0}-\xi_{0}e^{-\theta z_{0}}}\right\}
\end{align*}
and%
\begin{align*}
R_{m}  &  =\sum_{m=1}^{\infty}\frac{\,\left(  1-\theta\xi_{0}e^{-\theta z_{0}%
}\right)  ^{m}}{\left(  1-z_{0}-\xi_{0}e^{-\theta z_{0}}\right)  ^{m+1}}%
\frac{\left(  \zeta-z_{0}\right)  ^{m+1}}{\left(  m+1\right)  }-\sum
_{m=1}^{\infty}\frac{\,\left(  1-\theta\xi_{0}e^{-\theta z_{0}}\right)  ^{m}%
}{\left(  1-z_{0}-\xi_{0}e^{-\theta z_{0}}\right)  ^{m+1}}\frac{\left(
-z_{0}\right)  ^{m+1}}{\left(  m+1\right)  }\\
&  =\frac{1}{1-\theta\xi_{0}e^{-\theta z_{0}}}\left\{  -\ln\left(
\frac{\,1-z_{0}-\xi_{0}e^{-\theta z_{0}}-\left(  \zeta-z_{0}\right)  \left(
1-\theta\xi_{0}e^{-\theta z_{0}}\right)  }{1-z_{0}-\xi_{0}e^{-\theta z_{0}}%
}\right)  -\frac{\zeta\left(  1-\theta\xi_{0}e^{-\theta z_{0}}\right)
}{\left(  1-z_{0}-\xi_{0}e^{-\theta z_{0}}\right)  }\right\} \\
&  +\frac{1}{1-\theta\xi_{0}e^{-\theta z_{0}}}\ln\left(  \frac{\,1-z_{0}%
-\xi_{0}e^{-\theta z_{0}}+z_{0}\left(  1-\theta\xi_{0}e^{-\theta z_{0}%
}\right)  }{1-z_{0}-\xi_{0}e^{-\theta z_{0}}}\right) \\
&  =\frac{1}{1-\theta\xi_{0}e^{-\theta z_{0}}}\ln\left(  \frac{\,1-\xi
_{0}e^{-\theta z_{0}}\left(  1+z_{0}\theta\right)  }{1-\zeta-\xi_{0}e^{-\theta
z_{0}}\left(  1+\left(  z_{0}-\zeta\right)  \theta\right)  }\right)
-\frac{\zeta}{\left(  1-z_{0}-\xi_{0}e^{-\theta z_{0}}\right)  }%
\end{align*}
Hence,%
\begin{align}
t^{\ast}  &  =\frac{1}{1-\theta\xi_{0}e^{-\theta z_{0}}}\ln\left(  \frac
{1-\xi_{0}e^{-\theta z_{0}}\left(  1+z_{0}\theta\right)  }{1-\zeta-\xi
_{0}e^{-\theta z_{0}}\left(  1+\left(  z_{0}-\zeta\right)  \theta\right)
}\right)  +\nonumber\\
&  \sum_{m=1}^{\infty}\left[  \sum_{k=1}^{m-1}\frac{k!\,\sum_{j=0}^{k}%
\binom{m}{j}\left(  \xi_{0}e^{-\theta z_{0}}\right)  ^{k-j}\left(
-\theta\right)  ^{m-j}\mathcal{S}_{m-j}^{(k-j)}}{\left(  1-z_{0}-\xi
_{0}e^{-\theta z_{0}}\right)  ^{k+1}}\right]  \frac{\left(  \zeta
-z_{0}\right)  ^{m+1}-\left(  -z_{0}\right)  ^{m+1}}{\left(  m+1\right)  !}
\label{t*_with_log_term}%
\end{align}

Integration of the Taylor series in (\ref{Taylor_series_second_chc}),
$t^{\ast}=\int_{0}^{\zeta}\frac{dw}{1-w-\xi_{0}e^{-\theta w}}$ yields,
similarly as in (\ref{t*_with_log_term}),%
\begin{align*}
t^{\ast}  &  =\frac{1}{1-\theta\xi_{0}e^{-\theta z_{0}}}\ln\left(  \frac
{1-\xi_{0}e^{-\theta z_{0}}\left(  1+z_{0}\theta\right)  }{1-\zeta-\xi
_{0}e^{-\theta z_{0}}\left(  1+\left(  z_{0}-\zeta\right)  \theta\right)
}\right) \\
&  +\sum_{m=1}^{\infty}\left[  \sum_{k=1}^{m-1}\frac{\left(  \xi_{0}e^{-\theta
z_{0}}-\frac{1}{\theta}\right)  ^{k}\,\sum_{j=1}^{m-k}{\binom{k}{j}}\,\left(
\frac{\xi_{0}\theta e^{-\theta z_{0}}}{\theta\xi_{0}e^{-\theta z_{0}}%
-1}\right)  ^{j}j!T\left(  j,m-k\right)  }{\left(  1-z_{0}-\xi_{0}e^{-\theta
z_{0}}\right)  ^{k+1}}\right]  \,\left.  \frac{\left(  -\theta\right)
^{m}\left(  w-z_{0}\right)  ^{m+1}}{\left(  m+1\right)  }\right\vert
_{0}^{\zeta}%
\end{align*}
and (\ref{t*_with_log_term_chcster}). The series
(\ref{t*_with_log_term_chcster}) is numerically stabler than
(\ref{t*_with_log_term}), because all terms in the sums are positive.

The last sum can be rewritten as%
\begin{align*}
M  &  =\sum_{m=1}^{\infty}\left[  \sum_{k=1}^{m-1}\left(  \frac{\theta\xi
_{0}e^{-\theta z_{0}}-1}{\theta\left(  1-z_{0}-\xi_{0}e^{-\theta z_{0}%
}\right)  }\right)  ^{k}\sum_{j=1}^{m-k}{\binom{k}{j}}\,\left(  \frac{\xi
_{0}\theta e^{-\theta z_{0}}}{\theta\xi_{0}e^{-\theta z_{0}}-1}\right)
^{j}j!T\left(  j,m-k\right)  \right]  \,\frac{\theta^{m}z_{0}^{m+1}-\theta
^{m}\left(  z_{0}-\zeta\right)  ^{m+1}}{\left(  1-z_{0}-\xi_{0}e^{-\theta
z_{0}}\right)  \left(  m+1\right)  }\\
&  =\sum_{m=1}^{\infty}\left[  \sum_{l=1}^{m-1}\left(  \frac{\theta\xi
_{0}e^{-\theta z_{0}}-1}{\theta\left(  1-z_{0}-\xi_{0}e^{-\theta z_{0}%
}\right)  }\right)  ^{m-l}\sum_{j=1}^{l}{\binom{m-l}{j}}\,\left(  \frac
{\xi_{0}\theta e^{-\theta z_{0}}}{\theta\xi_{0}e^{-\theta z_{0}}-1}\right)
^{j}j!T\left(  j,l\right)  \right]  \,\frac{\theta^{m}z_{0}^{m+1}-\theta
^{m}\left(  z_{0}-\zeta\right)  ^{m+1}}{\left(  1-z_{0}-\xi_{0}e^{-\theta
z_{0}}\right)  \left(  m+1\right)  }%
\end{align*}
Reversing the $m$- and $l$-sum yields%
\begin{align*}
M  &  =\sum_{l=1}^{\infty}\sum_{m=l+1}^{\infty}\left(  \frac{\theta\xi
_{0}e^{-\theta z_{0}}-1}{\theta\left(  1-z_{0}-\xi_{0}e^{-\theta z_{0}%
}\right)  }\right)  ^{m-l}\sum_{j=1}^{l}{\binom{m-l}{j}}\,\left(  \frac
{\xi_{0}\theta e^{-\theta z_{0}}}{\theta\xi_{0}e^{-\theta z_{0}}-1}\right)
^{j}j!T\left(  j,l\right)  \,\frac{\theta^{m}z_{0}^{m+1}-\theta^{m}\left(
z_{0}-\zeta\right)  ^{m+1}}{\left(  1-z_{0}-\xi_{0}e^{-\theta z_{0}}\right)
\left(  m+1\right)  }\\
&  =\sum_{l=1}^{\infty}\sum_{j=1}^{l}\,\left(  \frac{\xi_{0}\theta e^{-\theta
z_{0}}}{\theta\xi_{0}e^{-\theta z_{0}}-1}\right)  ^{j}j!T\left(  j,l\right)
\sum_{m=1}^{\infty}{\binom{m}{j}}\left(  \frac{\theta\xi_{0}e^{-\theta z_{0}%
}-1}{1-z_{0}-\xi_{0}e^{-\theta z_{0}}}\right)  ^{m}\,\frac{\theta^{l}\left(
z_{0}^{m+l+1}-\left(  z_{0}-\zeta\right)  ^{m+l+1}\right)  }{\left(
1-z_{0}-\xi_{0}e^{-\theta z_{0}}\right)  \left(  m+l+1\right)  }%
\end{align*}
and%
\begin{align*}
M  &  =\sum_{l=1}^{\infty}\sum_{j=1}^{l}\,\left(  \frac{\xi_{0}\theta
e^{-\theta z_{0}}}{\theta\xi_{0}e^{-\theta z_{0}}-1}\right)  ^{j}T\left(
j,l\right)  \sum_{m=1}^{\infty}\frac{m!}{\left(  m-l\right)  !}\left(
\frac{\theta\xi_{0}e^{-\theta z_{0}}-1}{1-z_{0}-\xi_{0}e^{-\theta z_{0}}%
}\right)  ^{m}\,\frac{\theta^{l}\left(  z_{0}^{m+l+1}-\left(  z_{0}%
-\zeta\right)  ^{m+l+1}\right)  }{\left(  1-z_{0}-\xi_{0}e^{-\theta z_{0}%
}\right)  \left(  m+l+1\right)  }\\
&  =\frac{1}{1-z_{0}-\xi_{0}e^{-\theta z_{0}}}\sum_{l=1}^{\infty}\sum
_{j=1}^{l}\,\left(  \frac{\xi_{0}\theta e^{-\theta z_{0}}}{\theta\xi
_{0}e^{-\theta z_{0}}-1}\right)  ^{j}T\left(  j,l\right)  \left\{  \sum
_{m=1}^{\infty}\frac{\frac{m!}{\left(  m-l\right)  !}\left(  \frac{\theta
\xi_{0}e^{-\theta z_{0}}-1}{1-z_{0}-\xi_{0}e^{-\theta z_{0}}}z_{0}\right)
^{m}}{m+l+1}\right\}  \,\theta^{l}z_{0}^{l+1}\\
&  -\frac{1}{1-z_{0}-\xi_{0}e^{-\theta z_{0}}}\sum_{l=1}^{\infty}\sum
_{j=1}^{l}\,\left(  \frac{\xi_{0}\theta e^{-\theta z_{0}}}{\theta\xi
_{0}e^{-\theta z_{0}}-1}\right)  ^{j}T\left(  j,l\right)  \left\{  \sum
_{m=1}^{\infty}\frac{\frac{m!}{\left(  m-l\right)  !}\left(  \frac{\theta
\xi_{0}e^{-\theta z_{0}}-1}{1-z_{0}-\xi_{0}e^{-\theta z_{0}}}\left(
z_{0}-\zeta\right)  \right)  ^{m}}{m+l+1}\right\}  \,\theta^{l}\left(
z_{0}-\zeta\right)  ^{l+1}%
\end{align*}
The $m$-sum, which is of the type $\sum_{m=1}^{\infty}\frac{\frac{m!}{\left(
m-l\right)  !}x^{m}}{m+l+1}$, can be evaluated, because $\frac{m!}{\left(
m-l\right)  !\left(  m+l+1\right)  }$ is polynomial in $m$ plus $\frac{\alpha
}{m+l+1}$. The polynomials corresponds to derivatives of $\left(  1-x\right)
^{-q}$ and the $\frac{\alpha}{m+l+1}$ will generate a logarithm. Below we
compute the case terms up to $l=2$, but concentrated on
(\ref{t*_with_log_term_chcster}).

\subsection{Splitting off the $k=m-1$ term in
(\ref{Taylor_series_H(z)_rond_z0})}

We split-off the $k=m-1$ term in (\ref{t*_with_log_term_chcster}),%
\begin{align*}
t^{\ast}  &  =\frac{1}{1-\theta\xi_{0}e^{-\theta z_{0}}}\ln\left(  \frac
{1-\xi_{0}e^{-\theta z_{0}}\left(  1+z_{0}\theta\right)  }{1-\zeta-\xi
_{0}e^{-\theta z_{0}}\left(  1+\left(  z_{0}-\zeta\right)  \theta\right)
}\right) \\
&  +\frac{1}{2}\left(  \frac{\xi_{0}\theta^{2}e^{-\theta z_{0}}}{\left(
1-\theta\xi_{0}e^{-\theta z_{0}}\right)  ^{2}}\right)  \sum_{m=1}^{\infty
}\left(  \frac{\theta\xi_{0}e^{-\theta z_{0}}-1}{1-z_{0}-\xi_{0}e^{-\theta
z_{0}}}\right)  ^{m}\,\frac{\left(  m-1\right)  \left\{  z_{0}^{m+1}-\left(
z_{0}-\zeta\right)  ^{m+1}\right\}  }{\left(  m+1\right)  }\\
&  +\sum_{m=1}^{\infty}\left[  \sum_{k=1}^{m-2}\left(  \frac{\theta\xi
_{0}e^{-\theta z_{0}}-1}{\theta\left(  1-z_{0}-\xi_{0}e^{-\theta z_{0}%
}\right)  }\right)  ^{k}\sum_{j=1}^{m-k}{\binom{k}{j}}\,\left(  \frac{\xi
_{0}\theta e^{-\theta z_{0}}}{\theta\xi_{0}e^{-\theta z_{0}}-1}\right)
^{j}j!T\left(  j,m-k\right)  \right]  \,\frac{\theta^{m}z_{0}^{m+1}-\theta
^{m}\left(  z_{0}-\zeta\right)  ^{m+1}}{\left(  1-z_{0}-\xi_{0}e^{-\theta
z_{0}}\right)  \left(  m+1\right)  }%
\end{align*}
Now,%
\[
R_{m-1}=\frac{1}{2}\left(  \frac{\xi_{0}\theta^{2}e^{-\theta z_{0}}}{\left(
1-\theta\xi_{0}e^{-\theta z_{0}}\right)  ^{2}}\right)  \sum_{m=1}^{\infty
}\left(  \frac{\theta\xi_{0}e^{-\theta z_{0}}-1}{1-z_{0}-\xi_{0}e^{-\theta
z_{0}}}\right)  ^{m}\,\frac{m-1}{m+1}\left\{  z_{0}^{m+1}-\left(  z_{0}%
-\zeta\right)  ^{m+1}\right\}
\]
and with $\frac{m-1}{m+1}=1-\frac{2}{m+1}$, the sum becomes%
\begin{align*}
Q  &  =\sum_{m=1}^{\infty}\left(  \frac{\theta\xi_{0}e^{-\theta z_{0}}%
-1}{1-z_{0}-\xi_{0}e^{-\theta z_{0}}}\right)  ^{m}\,\frac{m-1}{m+1}\left\{
z_{0}^{m+1}-\left(  z_{0}-\zeta\right)  ^{m+1}\right\} \\
&  =\sum_{m=1}^{\infty}\left(  \frac{\theta\xi_{0}e^{-\theta z_{0}}-1}%
{1-z_{0}-\xi_{0}e^{-\theta z_{0}}}\right)  ^{m}\,\left\{  z_{0}^{m+1}-\left(
z_{0}-\zeta\right)  ^{m+1}\right\}  -2\sum_{m=1}^{\infty}\left(  \frac
{\theta\xi_{0}e^{-\theta z_{0}}-1}{1-z_{0}-\xi_{0}e^{-\theta z_{0}}}\right)
^{m}\,\left\{  \frac{z_{0}^{m+1}-\left(  z_{0}-\zeta\right)  ^{m+1}}%
{m+1}\right\} \\
&  =\frac{\theta\xi_{0}e^{-\theta z_{0}}-1}{1-z_{0}-\xi_{0}e^{-\theta z_{0}}%
}\left\{  \zeta_{0}^{2}\sum_{m=0}^{\infty}\left(  \frac{\theta\xi
_{0}e^{-\theta z_{0}}-1}{1-z_{0}-\xi_{0}e^{-\theta z_{0}}}z_{0}\right)
^{m}\,-\left(  z_{0}-\zeta\right)  ^{2}\sum_{m=0}^{\infty}\left(  \frac
{\theta\xi_{0}e^{-\theta z_{0}}-1}{1-z_{0}-\xi_{0}e^{-\theta z_{0}}}\left(
z_{0}-\zeta\right)  \right)  ^{m}\right\}  \,\\
&  -2\frac{1-z_{0}-\xi_{0}e^{-\theta z_{0}}}{\theta\xi_{0}e^{-\theta z_{0}}%
-1}\left\{  \sum_{m=1}^{\infty}\left(  \frac{\theta\xi_{0}e^{-\theta z_{0}}%
-1}{1-z_{0}-\xi_{0}e^{-\theta z_{0}}}\right)  ^{m}\,\frac{\left\{  z_{0}%
^{m}-\left(  z_{0}-\zeta\right)  ^{m}\right\}  }{m}-\frac{\theta\xi
_{0}e^{-\theta z_{0}}-1}{\left(  1-z_{0}-\xi_{0}e^{-\theta z_{0}}\right)
}\zeta\right\} \\
&  =\frac{\theta\xi_{0}e^{-\theta z_{0}}-1}{1-z_{0}-\xi_{0}e^{-\theta z_{0}}%
}\left\{  \frac{z_{0}^{2}}{1-\frac{\theta\xi_{0}e^{-\theta z_{0}}-1}{\left(
1-z_{0}-\xi_{0}e^{-\theta z_{0}}\right)  }z_{0}}\,-\frac{\left(  z_{0}%
-\zeta\right)  ^{2}}{1-\frac{\theta\xi_{0}e^{-\theta z_{0}}-1}{\left(
1-z_{0}-\xi_{0}e^{-\theta z_{0}}\right)  }\left(  z_{0}-\zeta\right)
}\right\} \\
&  -2\frac{1-z_{0}-\xi_{0}e^{-\theta z_{0}}}{\theta\xi_{0}e^{-\theta z_{0}}%
-1}\left\{  -\log\left(  \frac{1-\frac{\theta\xi_{0}e^{-\theta z_{0}}%
-1}{1-z_{0}-\xi_{0}e^{-\theta z_{0}}}z_{0}}{1-\frac{\theta\xi_{0}e^{-\theta
z_{0}}-1}{1-z_{0}-\xi_{0}e^{-\theta z_{0}}}\left(  z_{0}-\zeta\right)
}\right)  -\frac{\theta\xi_{0}e^{-\theta z_{0}}-1}{\left(  1-z_{0}-\xi
_{0}e^{-\theta z_{0}}\right)  }\zeta\,\right\}
\end{align*}
and%
\begin{align*}
Q  &  =\left(  \theta\xi_{0}e^{-\theta z_{0}}-1\right)  \left\{  \frac
{z_{0}^{2}}{1-\xi_{0}e^{-\theta z_{0}}\left(  1+z_{0}\theta\right)  }%
\,-\frac{\left(  z_{0}-\zeta\right)  ^{2}}{1-\zeta-\xi_{0}e^{-\theta z_{0}%
}\left(  1+\left(  z_{0}-\zeta\right)  \theta\right)  }\right\} \\
&  +2\frac{1-z_{0}-\xi_{0}e^{-\theta z_{0}}}{\theta\xi_{0}e^{-\theta z_{0}}%
-1}\log\left(  \frac{1-\xi_{0}e^{-\theta z_{0}}\left(  1+z_{0}\theta\right)
}{1-\zeta-\xi_{0}e^{-\theta z_{0}}\left(  1+\left(  z_{0}-\zeta\right)
\theta\right)  }\right)  +2\zeta\,
\end{align*}
Substituting $Q$ yields (\ref{t*_afgesplits_k=m-1}).

The form (\ref{t*_afgesplits_k=m-1}) is again better, but after summing the
last $m$-sum, we converge to the same results as in the Taylor series
(\ref{t*_with_log_term_chcster}) and (\ref{Taylor_series_H(z)_rond_z0}).

\subsection{Splitting off the $k=m-2$ term in
(\ref{Taylor_series_H(z)_rond_z0})}

We may continue in summing in this way. A next split-off in the $k$-sum for
$k=m-2$ is%
\[
R_{m-2}=\sum_{m=1}^{\infty}\left[  \left(  \frac{\theta\xi_{0}e^{-\theta
z_{0}}-1}{\theta\left(  1-z_{0}-\xi_{0}e^{-\theta z_{0}}\right)  }\right)
^{m-2}\sum_{j=1}^{2}{\binom{m-2}{j}}\,\left(  \frac{\xi_{0}\theta e^{-\theta
z_{0}}}{\theta\xi_{0}e^{-\theta z_{0}}-1}\right)  ^{j}j!T\left(  j,2\right)
\right]  \,\frac{\theta^{m}z_{0}^{m+1}-\theta^{m}\left(  z_{0}-\zeta\right)
^{m+1}}{\left(  1-z_{0}-\xi_{0}e^{-\theta z_{0}}\right)  \left(  m+1\right)  }%
\]
With $T\left(  1,2\right)  =\frac{1}{6}$ and $T\left(  2,2\right)  =\frac
{1}{8}$, we find%
\begin{align*}
R_{m-2}  &  =\sum_{m=1}^{\infty}\left(  \frac{\theta\xi_{0}e^{-\theta z_{0}%
}-1}{\theta\left(  1-z_{0}-\xi_{0}e^{-\theta z_{0}}\right)  }\right)
^{m-2}\frac{\left(  m-2\right)  }{6}\,\left(  \frac{\xi_{0}\theta e^{-\theta
z_{0}}}{\theta\xi_{0}e^{-\theta z_{0}}-1}\right)  \,\frac{\theta^{m}%
z_{0}^{m+1}-\theta^{m}\left(  z_{0}-\zeta\right)  ^{m+1}}{\left(  1-z_{0}%
-\xi_{0}e^{-\theta z_{0}}\right)  \left(  m+1\right)  }\\
&  +\sum_{m=1}^{\infty}\left(  \frac{\theta\xi_{0}e^{-\theta z_{0}}-1}%
{\theta\left(  1-z_{0}-\xi_{0}e^{-\theta z_{0}}\right)  }\right)  ^{m-2}%
\frac{\left(  m-2\right)  \left(  m-3\right)  }{8}\,\left(  \frac{\xi
_{0}\theta e^{-\theta z_{0}}}{\theta\xi_{0}e^{-\theta z_{0}}-1}\right)
^{2}\,\frac{\theta^{m}z_{0}^{m+1}-\theta^{m}\left(  z_{0}-\zeta\right)
^{m+1}}{\left(  1-z_{0}-\xi_{0}e^{-\theta z_{0}}\right)  \left(  m+1\right)
}\\
&  =\frac{\xi_{0}\theta^{3}e^{-\theta z_{0}}\left(  1-z_{0}-\xi_{0}e^{-\theta
z_{0}}\right)  }{6\left(  \theta\xi_{0}e^{-\theta z_{0}}-1\right)  ^{3}}%
\sum_{m=1}^{\infty}\left(  \frac{\theta\xi_{0}e^{-\theta z_{0}}-1}{1-z_{0}%
-\xi_{0}e^{-\theta z_{0}}}\right)  ^{m}\,\frac{\left(  m-2\right)  \left(
z_{0}^{m+1}-\left(  z_{0}-\zeta\right)  ^{m+1}\right)  }{\left(  m+1\right)
}\\
&  +\,\,\frac{\xi_{0}^{2}\theta^{4}e^{-2\theta z_{0}}\left(  1-z_{0}-\xi
_{0}e^{-\theta z_{0}}\right)  }{8\left(  \theta\xi_{0}e^{-\theta z_{0}%
}-1\right)  ^{4}}\sum_{m=1}^{\infty}\left(  \frac{\theta\xi_{0}e^{-\theta
z_{0}}-1}{1-z_{0}-\xi_{0}e^{-\theta z_{0}}}\right)  ^{m}\frac{\left(
m-2\right)  \left(  m-3\right)  \left(  z_{0}^{m+1}-\left(  z_{0}%
-\zeta\right)  ^{m+1}\right)  }{\left(  m+1\right)  }%
\end{align*}
We recognize that the first series is similar to $Q$, because $\frac{m-2}%
{m+1}=1-\frac{3}{m+1}$ and thus equal to%
\begin{align*}
Q_{\ast}  &  =\left(  \theta\xi_{0}e^{-\theta z_{0}}-1\right)  \left\{
\frac{z_{0}^{2}}{1-\xi_{0}e^{-\theta z_{0}}\left(  1+z_{0}\theta\right)
}\,-\frac{\left(  z_{0}-\zeta\right)  ^{2}}{1-\zeta-\xi_{0}e^{-\theta z_{0}%
}\left(  1+\left(  z_{0}-\zeta\right)  \theta\right)  }\right\} \\
&  +3\frac{1-z_{0}-\xi_{0}e^{-\theta z_{0}}}{\theta\xi_{0}e^{-\theta z_{0}}%
-1}\log\left(  \frac{1-\xi_{0}e^{-\theta z_{0}}\left(  1+z_{0}\theta\right)
}{1-\zeta-\xi_{0}e^{-\theta z_{0}}\left(  1+\left(  z_{0}-\zeta\right)
\theta\right)  }\right)  +3\zeta\,
\end{align*}
while, with $\frac{\left(  m-2\right)  \left(  m-3\right)  }{m+1}=\left(
m-6\right)  +\frac{12}{m+1}=m-6\left(  1-\frac{2}{m+1}\right)  $, the last sum
contains precisely $Q$ and a new series%
\begin{align*}
R_{m-2}  &  =\frac{\xi_{0}\theta^{3}e^{-\theta z_{0}}\left(  1-z_{0}-\xi
_{0}e^{-\theta z_{0}}\right)  }{6\left(  \theta\xi_{0}e^{-\theta z_{0}%
}-1\right)  ^{2}}\left\{  \frac{z_{0}^{2}}{1-\xi_{0}e^{-\theta z_{0}}\left(
1+z_{0}\theta\right)  }\,-\frac{\left(  z_{0}-\zeta\right)  ^{2}}{1-\zeta
-\xi_{0}e^{-\theta z_{0}}\left(  1+\left(  z_{0}-\zeta\right)  \theta\right)
}\right\} \\
&  +\frac{\xi_{0}\theta^{3}e^{-\theta z_{0}}\left(  1-z_{0}-\xi_{0}e^{-\theta
z_{0}}\right)  ^{2}}{2\left(  \theta\xi_{0}e^{-\theta z_{0}}-1\right)  ^{3}%
}\frac{1}{\theta\xi_{0}e^{-\theta z_{0}}-1}\log\left(  \frac{1-\xi
_{0}e^{-\theta z_{0}}\left(  1+z_{0}\theta\right)  }{1-\zeta-\xi_{0}e^{-\theta
z_{0}}\left(  1+\left(  z_{0}-\zeta\right)  \theta\right)  }\right) \\
&  +\frac{\xi_{0}\theta^{3}e^{-\theta z_{0}}\left(  1-z_{0}-\xi_{0}e^{-\theta
z_{0}}\right)  }{2\left(  \theta\xi_{0}e^{-\theta z_{0}}-1\right)  ^{3}}%
\zeta\,\\
&  +\frac{\xi_{0}^{2}\theta^{4}e^{-2\theta z_{0}}\left(  1-z_{0}-\xi
_{0}e^{-\theta z_{0}}\right)  }{8\left(  \theta\xi_{0}e^{-\theta z_{0}%
}-1\right)  ^{4}}\sum_{m=1}^{\infty}\left(  \frac{\theta\xi_{0}e^{-\theta
z_{0}}-1}{1-z_{0}-\xi_{0}e^{-\theta z_{0}}}\right)  ^{m}m\left(  z_{0}%
^{m+1}-\left(  z_{0}-\zeta\right)  ^{m+1}\right) \\
&  -6\frac{\xi_{0}^{2}\theta^{4}e^{-2\theta z_{0}}\left(  1-z_{0}-\xi
_{0}e^{-\theta z_{0}}\right)  }{8\left(  \theta\xi_{0}e^{-\theta z_{0}%
}-1\right)  ^{4}}Q
\end{align*}
The new series
\[
W=\sum_{m=1}^{\infty}\left(  \frac{\theta\xi_{0}e^{-\theta z_{0}}-1}%
{1-z_{0}-\xi_{0}e^{-\theta z_{0}}}\right)  ^{m}m\left(  z_{0}^{m+1}-\left(
z_{0}-\zeta\right)  ^{m+1}\right)
\]
follows from $\frac{1}{\left(  1-x\right)  ^{2}}=\frac{d}{dx}\left(  \frac
{1}{1-x}\right)  =\sum_{m=1}^{\infty}mx^{m-1}$ as%
\[
W=\left(  \theta\xi_{0}e^{-\theta z_{0}}-1\right)  \left(  1-z_{0}-\xi
_{0}e^{-\theta z_{0}}\right)  \left\{  \frac{z_{0}^{2}}{\left(  1-\xi
_{0}e^{-\theta z_{0}}\left(  1+z_{0}\theta\right)  \right)  ^{2}}%
-\frac{\left(  z_{0}-\zeta\right)  ^{2}}{\left(  1-\zeta-\xi_{0}e^{-\theta
z_{0}}\left(  1+\left(  z_{0}-\zeta\right)  \theta\right)  \right)  ^{2}%
}\right\}
\]
Hence,%
\begin{align*}
R_{m-2}  &  =\frac{\xi_{0}\theta^{3}e^{-\theta\zeta_{0}}\left(  1-\zeta
_{0}-\xi_{0}e^{-\theta\zeta_{0}}\right)  }{6\left(  1-\theta\xi_{0}%
e^{-\theta\zeta_{0}}\right)  ^{2}}\left\{  \frac{\zeta_{0}^{2}}{1-\xi
_{0}e^{-\theta\zeta_{0}}\left(  1+\zeta_{0}\theta\right)  }\,-\frac{\left(
\zeta_{0}-\zeta\right)  ^{2}}{1-\zeta-\xi_{0}e^{-\theta\zeta_{0}}\left(
1+\left(  \zeta_{0}-\zeta\right)  \theta\right)  }\right\} \\
&  +\frac{\xi_{0}\theta^{3}e^{-\theta\zeta_{0}}\left(  1-\zeta_{0}-\xi
_{0}e^{-\theta\zeta_{0}}\right)  ^{2}}{2\left(  1-\theta\xi_{0}e^{-\theta
\zeta_{0}}\right)  ^{3}}\frac{1}{1-\theta\xi_{0}e^{-\theta\zeta_{0}}}%
\log\left(  \frac{1-\xi_{0}e^{-\theta\zeta_{0}}\left(  1+\zeta_{0}%
\theta\right)  }{1-\zeta-\xi_{0}e^{-\theta\zeta_{0}}\left(  1+\left(
\zeta_{0}-\zeta\right)  \theta\right)  }\right) \\
&  -\frac{\xi_{0}\theta^{3}e^{-\theta\zeta_{0}}\left(  1-\zeta_{0}-\xi
_{0}e^{-\theta\zeta_{0}}\right)  }{2\left(  1-\theta\xi_{0}e^{-\theta\zeta
_{0}}\right)  ^{3}}\zeta\,\\
&  -\frac{\xi_{0}^{2}\theta^{4}e^{-2\theta\zeta_{0}}\left(  1-\zeta_{0}%
-\xi_{0}e^{-\theta\zeta_{0}}\right)  ^{2}}{8\left(  1-\theta\xi_{0}%
e^{-\theta\zeta_{0}}\right)  ^{3}}\left\{  \frac{\zeta_{0}^{2}}{\left(
1-\xi_{0}e^{-\theta\zeta_{0}}\left(  1+\theta\zeta_{0}\right)  \right)  ^{2}%
}-\frac{\left(  \zeta_{0}-\zeta\right)  ^{2}}{\left(  1-\zeta-\xi
_{0}e^{-\theta\zeta_{0}}\left(  1+\theta\left(  \zeta_{0}-\zeta\right)
\right)  \right)  ^{2}}\right\} \\
&  +3\frac{\xi_{0}^{2}\theta^{4}e^{-2\theta\zeta_{0}}\left(  1-\zeta_{0}%
-\xi_{0}e^{-\theta\zeta_{0}}\right)  }{4\left(  1-\theta\xi_{0}e^{-\theta
\zeta_{0}}\right)  ^{3}}\left\{  \frac{\zeta_{0}^{2}}{1-\xi_{0}e^{-\theta
\zeta_{0}}\left(  1+\zeta_{0}\theta\right)  }\,-\frac{\left(  \zeta_{0}%
-\zeta\right)  ^{2}}{1-\zeta-\xi_{0}e^{-\theta\zeta_{0}}\left(  1+\left(
\zeta_{0}-\zeta\right)  \theta\right)  }\right\} \\
&  +3\frac{\xi_{0}^{2}\theta^{4}e^{-2\theta\zeta_{0}}\left(  1-\zeta_{0}%
-\xi_{0}e^{-\theta\zeta_{0}}\right)  ^{2}}{2\left(  1-\theta\xi_{0}%
e^{-\theta\zeta_{0}}\right)  ^{4}}\left\{  \frac{1}{1-\theta\xi_{0}%
e^{-\theta\zeta_{0}}}\log\left(  \frac{1-\xi_{0}e^{-\theta\zeta_{0}}\left(
1+\zeta_{0}\theta\right)  }{1-\zeta-\xi_{0}e^{-\theta\zeta_{0}}\left(
1+\left(  \zeta_{0}-\zeta\right)  \theta\right)  }\right)  \right\} \\
&  -3\frac{\zeta\xi_{0}^{2}\theta^{4}e^{-2\theta\zeta_{0}}\left(  1-\zeta
_{0}-\xi_{0}e^{-\theta\zeta_{0}}\right)  }{2\left(  1-\theta\xi_{0}%
e^{-\theta\zeta_{0}}\right)  ^{4}}%
\end{align*}
Collecting all results in (\ref{t*_afgesplits_k=m-2}). The last sum in
(\ref{t*_afgesplits_k=m-2}) is small and $O\left(  \zeta_{0}^{5}-\left(
\zeta_{0}-\zeta\right)  ^{5}\right)  $ and only plays a role when
$\zeta\rightarrow\zeta_{\max}$. Also, smaller $\theta$ result in faster
convergence (only checked for $\zeta_{0}=\frac{\zeta}{2}$). In summary, we
have shown that, to any desired accuracy, the integral
(\ref{Kermack_McKendrick_integral_SIR_removed_fractions}) can be analytically
approximated. Moreover, ignoring the remaining $m$-sum, all analytic terms
lower bound the integral
(\ref{Kermack_McKendrick_integral_SIR_removed_fractions}).

\section{Coefficients $a_{k}\left(  m,j\right)  $ of the polynomial $p\left(
x;m,j\right)  $ in (\ref{def_pol_zeta_m(t_0)})}

\label{sec_coefficients_ak(m,j)}We revisit and rewrite the form
(\ref{Taylor_coefficient_zeta_rond_t0}) as%
\[
\zeta_{m}\left(  t_{0}^{\ast}\right)  =\frac{\left(  -1\right)  ^{m-1}Z}%
{m!}-\frac{\left(  -1\right)  ^{m}}{m!\theta}\sum_{j=1}^{m}\left(
-A\theta\right)  ^{j}\sum_{k=0}^{m-j}a_{k}\left(  m,j\right)  x^{k}%
\]
where $p\left(  x;m,j\right)  =\sum_{k=0}^{m-j}a_{k}\left(  m,j\right)  x^{k}$
reduces for $p\left(  x;m,m\right)  =a_{0}\left(  m,m\right)  =\left(
m-1\right)  !$. The first order polynomial $p\left(  x;m,m-1\right)
=a_{0}\left(  m,m-1\right)  +a_{1}\left(  m,m-1\right)  x$ for $m\geq2$ and we
list the coefficients for a few $m$,%
\[%
\begin{tabular}
[c]{ccc}%
$m$ & $a_{0}\left(  m,m-1\right)  $ & $a_{1}\left(  m,m-1\right)  $\\
2 & 1 & 1\\
3 & 2 & 3\\
4 & 7 & 12\\
5 & 33 & 60\\
6 & 192 & 360\\
7 & 1320 & 2 520\\
8 & 10440 & 20160\\
9 & 93240 & 181440\\
10 & 927360 & 1814400
\end{tabular}
\
\]
By inspection, we deduce that $a_{1}\left(  m,m-1\right)  =\frac{m!}{2}$ and
$a_{0}\left(  m,m-1\right)  =\frac{m!}{4}+\frac{\left(  m-2\right)  !}{2}$.

The second order polynomial $p\left(  x;m,m-2\right)  =a_{0}\left(
m,m-2\right)  +a_{1}\left(  m,m-2\right)  x+a_{2}\left(  m,m-2\right)  x^{2}$
has coefficients%

\[%
\begin{tabular}
[c]{cccc}%
$m$ & $a_{0}\left(  m,m-2\right)  $ & $a_{1}\left(  m,m-2\right)  $ &
$a_{2}\left(  m,m-2\right)  $\\
3 & 1 & 2 & 1\\
4 & 3 & 11 & 7\\
5 & 17 & 69 & 50\\
6 & 120 & 499 & 390\\
7 & 979 & 4096 & 3360\\
8 & 8991 & 37640 & 31920\\
9 & 91586 & 382920 & 332640\\
10 & 1024022 & 4273080 & 3780000
\end{tabular}
\
\]
By inspection, we obtain $a_{2}\left(  m,m-2\right)  =\frac{m!(3m-5)}{24}$
and
\[
a_{1}\left(  m,m-2\right)  =\frac{\left(  m-3\right)  !}{72}\left(
24-68m+57m^{2}-34m^{3}+9m^{4}\right)
\]
The latter is found as solution of a difference equation
\[
\frac{3!a_{1}\left(  m,m-2\right)  }{\left(  m-3\right)  !}-\frac
{3!a_{1}\left(  m,m-3\right)  }{\left(  m-4\right)  !}=3(m-1)^{2}-\left(
m-2\right)  \left(  m-4\right)
\]
leading to a summation of the right-hand side\footnote{Summations of powers of
integers can be expressed as Bernoulli polynomials \cite{Rademacher}.}.
However, $a_{0}\left(  m,m-2\right)  $ possesses a more complicated law, which
has defeated us so far.

The highest order polynomial $p\left(  x;m,1\right)  =\sum_{k=0}^{m-1}%
a_{k}\left(  m,1\right)  x^{k}$ has $a_{0}\left(  m,1\right)  =a_{m-1}\left(
m,1\right)  =1$ and $a_{1}\left(  m,1\right)  =m-1$. The coefficient
$a_{m-2}\left(  m,1\right)  =\frac{\left(  m-1\right)  \left(  m-2\right)
}{2}+1$.

\end{document}